\documentclass[12pt,hidelinks]{article}
\pdfoutput=1
\usepackage[body={17.5cm, 22cm},right=2cm]{geometry}
\usepackage[table]{xcolor}
\usepackage{color}
\usepackage{amssymb,graphicx}
\usepackage{amsmath}
\usepackage{epsfig}
\usepackage{lmodern}
\usepackage[normalem]{ulem}
\usepackage{bm}
\usepackage[bookmarks=true,pdfborder={0 0 0}]{hyperref} 
\usepackage{feynmp-auto}
\usepackage{tabularx}
\usepackage{slashed}
\usepackage{amsthm}

\usepackage{authblk}

\usepackage{caption}
\usepackage{subcaption}

\usepackage{cite} 
\usepackage{hyperref}

\newtheorem*{remark}{Claim}

\def\beq{\begin{equation}}
\def\eeq{\end{equation}}
\def\bea{\begin{eqnarray}}
\def\eea{\end{eqnarray}}
\def\bitem{\begin{itemize}}
	\def\eitem{\end{itemize}}
\newcommand{\bec}{\begin{center}}
	\newcommand{\eec}{\end{center}}
\newcommand{\ba}{\begin{array}}
	\newcommand{\ea}{\end{array}}

\usepackage{float}
\allowdisplaybreaks

\definecolor{myred}{RGB}{144,10,2}

\definecolor{mygreen}{RGB}{60,200,50}

\begin{document}

\setcounter{page}{0}
\thispagestyle{empty}

\parskip 3pt

\font\mini=cmr10 at 2pt

\begingroup
\renewcommand{\thefootnote}{\alph{footnote}} 
\setcounter{footnote}{0} 

\begin{titlepage}
\noindent \makebox[15cm][l]{\footnotesize \hspace*{-.2cm} }  \\  [-1mm]
~\vspace{2cm}

\begin{center}

{\LARGE \bf  Metastable cosmic strings are broken
\\ \vspace{0.4cm}
 at the start}

\vspace{0.9cm}

{\large
Lorenzo Tranchedone\,\footnote{\url{lorenzo.tranchedone@physics.ox.ac.uk}}, Ethan Carragher, Edward Hardy,\\ \vspace{0.2cm} Natálie Koscelanská van IJcken} 
\\
\vspace{.6cm}
{\normalsize { \sl Rudolf Peierls Centre for Theoretical Physics, University of Oxford, \\ Parks Road, Oxford OX1 3PU, UK}}

\end{center}

\vspace{.8cm}
\begin{abstract}

We show that metastable cosmic strings break at early times, either via finite-temperature effects or by attaching to pre-existing monopoles during network percolation. The resulting segments can be initially super-horizon in size and thus persist for a significant amount of time. If the strings do not re-percolate, the network’s eventual destruction is typically due to this early-time breaking rather than late-time quantum tunnelling. Survival of strings to epochs probed by NANOGrav requires $m_M^2/\mu\gtrsim 10^3$, where $m_M$ and $\mu$ are the monopole mass and the string tension respectively, over an order of magnitude larger than previous estimates. We also revisit quantum-tunnelling induced breaking. Results from numerical simulations suggest that this occurs mainly at rare high-tension points on the strings, yielding a rate much larger than is usually assumed. We briefly discuss the related scenario of flux tubes in a dark QCD-like hidden sector with dark-quark masses above the confinement scale.

\end{abstract}

\end{titlepage}

\endgroup 

\setcounter{footnote}{0}
\renewcommand{\thefootnote}{\arabic{footnote}}

{\fontsize{11}{11}
\tableofcontents
}

\newpage
\section{Introduction}

The possibility of cosmic strings in the early Universe is theoretically and observationally interesting. Many high-energy extensions of the Standard Model admit topologically stable string-like defects. Moreover, a network of strings automatically forms if the Hubble scale during inflation, or the post-inflationary reheating temperature, exceeds the associated symmetry breaking scale. Such networks can lead to a variety of distinctive signals. Most notably for our purposes, they produce a stochastic gravitational wave (GW) background with a spectrum that is approximately flat over a wide frequency range, enabling cross-correlation between multiple GW observatories. Because the network is long-lived, it connects late-time observations to physics at ultra-high energy scales.  Moreover, the strength of signals grows with the symmetry breaking scale, opposite to the trend of most other beyond Standard Model physics scenarios.

Cosmic strings are often metastable rather than absolutely stable.  
We focus on local cosmic strings arising from spontaneous breaking of a gauge symmetry, which are metastable if the breaking occurs in two stages, 
\beq \label{eq:2stage}
G\rightarrow H \rightarrow K~,
\eeq
with the first homotopy group $\pi_1(G/K)$ trivial but $\pi_1(H/K)$ not. 
The strings break by the spontaneous nucleation of a monopole-antimonopole pair \cite{Preskill:1992ck}, but they can be extremely long-lived if the two symmetry breaking scales are hierarchically separated. A simple example of such a breaking pattern is $\mathrm{SU}(2)\rightarrow \mathrm{U}(1) \rightarrow {\rm nothing}$, with the first breaking by an adjoint scalar, leading to `t Hooft–Polyakov monopoles. Similar patterns arise naturally in certain grand unified theories \cite{Kibble:1982ae,Jeannerot:2003qv,King:2021gmj,Dunsky:2021tih,Buchmuller:2023aus,Afzal:2023kqs,Lazarides:2023iim,Fu:2023mdu,Lazarides:2023rqf,Ahmed:2023pjl,Maji:2025thf} (see also 
\cite{Buchmuller:2021dtt,Afzal:2023cyp,Pallis:2024joc,Chitose:2025cmt,Antusch:2025xrs} for other high-energy scenarios that lead to metastable strings).

If a network is composed of metastable strings, then it inevitably decays away at some cosmological time, which can lead to distinctive observational imprints. These include a low-frequency cutoff in the GW power spectrum (set by the time of the network's destruction), the absence of signals in the cosmic microwave background despite earlier GW emission, or characteristic GW bursts from the motion of string endpoints during decay. Such a metastable network has been proposed as an explanation of the tentative NANOGrav ${\rm nHz}$ band GW signal \cite{NANOGrav:2023detection}, and it would provide a remarkable window to physics at extremely high energy scales.

It is therefore important to understand the cosmological evolution of metastable string networks in detail. The conventional picture is that such a network forms and initially evolves as if stable, approaching an approximately scale-invariant attractor with roughly one long string per Hubble patch. It decays once the breaking rate per string length $\Gamma$ is fast compared to Hubble, i.e. $H^2\simeq\Gamma$. The breaking by nucleation of monopole-antimonopole pairs through quantum tunnelling has been estimated to occur at a rate
\beq \label{eq:4}
\Gamma \simeq \frac{\mu}{2\pi} e^{-\pi \kappa}~, \quad  \kappa \equiv m_M^2/\mu ~,
\eeq
where $\mu$ is the string tension (its energy per unit length), and $m_M$ is the monopole mass.  If the network decays during radiation domination, the temperature of the Universe when this happens is related to the monopole mass by
\beq \label{eq:TdestQ}
T_d \simeq \sqrt{M_{\rm Pl}\mu^{1/2}} e^{-\pi m_M^2/(4\mu)}~,
\eeq
up to numerical coefficients of order unity in the prefactor, where $M_{\rm Pl}=1/G_{\rm N}^{1/2}$ is the Planck mass with $G_{\rm N}$ Newton's constant. Assuming Eq.~\eqref{eq:TdestQ} holds, the NANOGrav signal can be fit with values of $m_M^2/\mu\simeq 60$, such that $T_d \sim {\rm MeV}$.

In this paper, we show that across a broad class of formation scenarios the strings in a metastable network are already broken soon after the network forms, although the segments can be very super-horizon for $m_M/\sqrt{\mu}\gg 1$. 
The physical processes responsible for this early breaking depend on how the symmetry associated with the strings is restored. We analyse finite-temperature symmetry restoration and inflationary fluctuations, which we regard as the most generic mechanisms.\footnote{\label{fn:1} Other possibilities that we do not consider include symmetry breaking induced by a direct coupling between the inflaton and a scalar responsible for spontaneous symmetry breaking. We expect that in some models of this type, late-time breaking might dominate, in which case the values of $\kappa$ used in previous studies would be appropriate.} In both cases the breaking occurs at a rate $\mathcal{O}\left(e^{-c m_M/\sqrt{\mu}}\right)$, where $c$ is an order-one number, in contrast to the zero-temperature expectation of Eq.~\eqref{eq:4}.

We argue that subsequent interactions between long string segments do not regenerate infinite strings. If loops likewise do not combine to become arbitrarily large, which we suspect is unlikely especially after finite-temperature restoration, then early-time breaking is responsible for the network's ultimate destruction. This occurs when the finite segments, as well as the typically smaller loops, re-enter the horizon (these dynamics are reminiscent of the scenario considered in Ref.~\cite{Lazarides:2022jgr} in which a population of monopoles is partly inflated away). The resulting observational signals resemble those discussed in previous work, although achieving them requires much larger hierarchies between the monopole mass and the string tension for the formation scenarios we consider. 
We emphasise, however, that we have not proved that loops do not re-percolate, and clarifying their evolution is an important direction for future work.

We also revisit the late-time breaking rate of local strings by quantum tunnelling, taking into account that the string tension is not uniform in a realistic network. 
We find that localised regions of strings can have energy densities exceeding the average by factors of a few, for instance due to features such as kinks or cusps. Using numerical simulations, we estimate the frequency of such high-tension regions. Although relatively rare, the dependence of Eq.~\eqref{eq:4} on $\mu$ is so strong that, for the values $m_M^2/\mu$ of phenomenological interest, breaking at these points dominates. Consequently, even at zero temperature, Eq.~\eqref{eq:4} underestimates the breaking rate by many orders of magnitude.

Although we focus on local strings, the flux tubes of a new asymptotically free gauge group in a hidden sector can behave as metastable cosmic strings if the confinement scale lies well below the mass of the lightest hidden-sector ``quarks''. Given that hidden gauge sectors appear to be common in string theory compactifications (see, e.g., \cite{Giedt:2000bi,Cvetic:2004ui,Taylor:2015ppa,Acharya:2016fge,Acharya:2018deu}), this is a particularly well-motivated scenario, and we therefore comment on the similarities with our main results.

The structure of this paper is as follows. We analyse string networks formed  by finite-temperature effects in Section~\ref{sec:reheat} and by inflationary fluctuations in Section~\ref{sec:inflation}. We study the subsequent evolution of the resulting string segments in Section~\ref{sec:subsequent}, and in  Section~\ref{sec:late} we revisit the late-time breaking rate. The phenomenological and observational implications are discussed in Section~\ref{sec:signals}.
The case of flux tubes with heavy quarks is briefly discussed in Section~\ref{sec:qcdlike}. 
Section~\ref{sec:discuss} summarises our results and outlines directions for future work. 
We present our analytic argument that interactions between long strings do not lead to re-percolation in Appendix~\ref{app:toymodel}.

\section{Networks formed by finite temperature} \label{sec:reheat} 

We consider theories with the two-stage symmetry breaking pattern of Eq.~\eqref{eq:2stage}. One way a string network can form is if the maximum temperature of the Universe after inflation exceeds the associated symmetry breaking scale $v$. Finite-temperature effects then generically restore the symmetry and a network of strings forms when the temperature of the Universe drops below $v$. 
For simplicity, we phrase our discussion in terms of a symmetry that is broken by a complex scalar getting a vacuum expectation value, although the dynamics are similar in more complicated theories.

If the Hubble scale during inflation $H_I$ is relatively large, near the maximum allowed by searches for tensor modes ($H_I\lesssim 6\cdot 10^{13}\,{\rm GeV} $), and reheating is fast, the maximum temperature post-inflation might be as large as $10^{15}\,{\rm GeV}$, although this depends on the details of thermalisation. Such temperatures allow the formation of string networks with relatively large tensions (larger, in particular, than those possible from inflationary fluctuations), and correspondingly strong observational signals. 

When the temperature drops below $v$, the symmetry-breaking field is random on scales set by a correlation length $\xi$. 
This yields proto-string segments in regions of space with non-trivial field-space winding. 
Such segments connect together in a process resembling a random walk. 
In three spatial dimensions, this produces both closed loops and, in the absence of monopoles, infinite strings, since such a random walk  has an order-one probability of not returning to its starting point. The correlation length depends on the details of the phase transition (e.g., whether it is first order or a crossover), but on general grounds $T_\star^{-1} \lesssim \xi \lesssim H_\star^{-1}$ where $T_\star$ and $H_\star$ are the temperature and Hubble parameter at the transition. We consider theories without any exponentially small parameters, so $T_\star\simeq v$ and $H_\star \simeq v^2/M_{\rm Pl}$. 
For plots and where definiteness is useful, we fix $\mu=2\pi v^2$ where $v$ is the symmetry breaking scale associated with the strings (as for Abelian-Higgs theories at critical coupling). The symmetry-breaking phase transition associated with the monopoles is similarly assumed to occur at a temperature of order the monopole mass $m_M$.

Throughout our work, we consider a simplified scenario in which the Universe is instantaneously reheated to a temperature $T_{\rm RH}$, immediately thermalises, and enters radiation domination. In this section, we further assume $v\lesssim T_{\rm RH}\lesssim  m_M$. 
If the reheating temperature were higher, a large monopole abundance would be produced, and the random walk of proto-string segments would encounter one almost immediately, yielding only sub-horizon segments that would quickly decay. 

Moreover, we take the reheating temperature to be less than
the size of the monopoles, such that the produced monopoles can be treated as point-like particles. This assumption can also be viewed as a condition on the gauge coupling to not be too small.

Even for $T_{\rm RH}\ll m_M$, there is a thermally produced population of monopoles, albeit exponentially suppressed. These rare monopoles eventually terminate the proto–string random walks, and we study their impact in  Section~\ref{ss:thermalM}. 
Before this, we analyse the breaking of long strings shortly after their formation through the spontaneous nucleation of a monopole–antimonopole pair at finite temperature.

\subsection{String breaking at finite temperature}
\label{sec:stringbreaking}

Just as finite-temperature effects modify bubble nucleation in a first-order phase transition, they also alter the breaking rate of a metastable string, which proceeds via the nucleation of a monopole-antimonopole pair. Finite-temperature effects enhance the nucleation rate relative to the zero-temperature prediction of Eq.~\eqref{eq:4}, although the enhancement is substantial only at early times when the temperature is high.

At temperature $T$, the corrected version of Eq.~\eqref{eq:4} can be written in the generic form
\begin{equation}\label{eq:decayT}
    \Gamma(T) = C(T) \,e^{-S_B(T)} \, .
\end{equation}
Thermal effects can be separated into contributions to the prefactor $C(T)$ and the bounce action $S_B(T)$. The physical origins and importance of these depend on the temperature regime under consideration.

We recall that at zero temperature, the tunnelling rate is obtained from the well-known bounce action calculation on the Euclidean string worldsheet (see, e.g., \cite{Preskill:1992ck, Coleman:1978ae}). Neglecting the finite string thickness and monopole size,\footnote{This assumption was recently re-examined in \cite{Chitose_2024}, where the authors considered a specific realisation of an $\mathrm{SU}(2) \rightarrow \mathrm{U}(1) \rightarrow \mathrm{nothing}$ breaking pattern. For small scale separation (roughly $\sqrt{\kappa} \lesssim 10$) the bounce action was found to be somewhat larger than the analytic result obtained in the ``thin-string'' approximation. In the discussion below, we find that the phenomenologically interesting regime is $\sqrt{\kappa} \gg 10$, so we safely employ the thin-string approximation (at least as long as the temperature is not too large, see below).} the single instanton solution corresponding to monopole-antimonopole nucleation is a circular region of true vacuum (i.e., no string) of radius $r_M = m_M/\mu$.  Summing over all $n$-instanton configurations and integrating over the worldsheet yields 
\begin{equation}
    S_B(T=0) = \frac{\pi m_M^2}{\mu} \, , \quad C(T = 0) = \frac{\mu}{2\pi} \, ,
\end{equation}
which corresponds to Eq.~\eqref{eq:4} (analysis of the related process of a string decaying to a lower tension string can be found in \cite{Monin:2008mp}). 

Finite-temperature effects can be incorporated by compactifying the Euclidean time direction on a circle of period $\beta=1/T$, such that the instanton solutions become periodic in Euclidean time. Formally, we can think of these solutions as a tool to evaluate an integral along a Keldysh-Schwinger contour, see \cite{Steingasser:2023gde}. Finite $T$ effects have a major impact when $\beta$ is comparable to or smaller than the size of the vacuum instanton, $2r_M$. The critical temperature at which this occurs is
\begin{equation}
    T_0 = \frac{\mu}{2m_M} = \frac{m_M}{2 \kappa} = \frac{1}{2} \sqrt{\frac{\mu}{\kappa}}\, ,
\end{equation}
where, as in Eq.~\eqref{eq:4},
\beq
\kappa\equiv \frac{m_M^2}{\mu}~.
\eeq
When the temperature is finite but below $T_0$, Goldstone bosons (corresponding to transverse oscillations) that are thermally excited along the string provide the dominant modification to the breaking rate due to temperature. 
The propagation and collision of these excitations produce fluctuations in the distribution of energy in the string, which catalyse the decay. This effect was studied in Refs.~\cite{Monin:2008uj,Monin:2009ch,Monin:2009gi}, where the leading order correction to the prefactor was found to scale as $ (T/{T_0})^8$, while the exponential is unaffected. This contribution is therefore unimportant for our purposes and we neglect it in what follows.
\begin{figure}[t]
    \centering
    \begin{subfigure}[b]{0.32\textwidth}
         \centering
         \includegraphics[width=\textwidth]{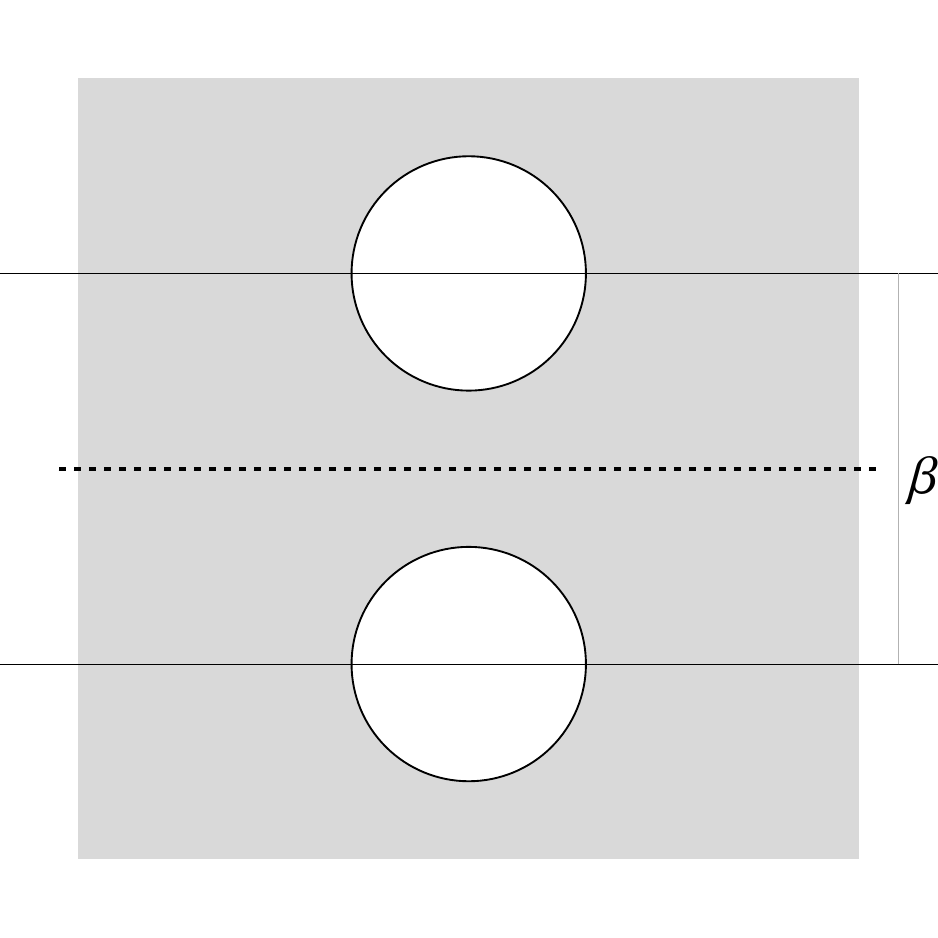}
         \subcaption{$T< T_0$}
    \end{subfigure}\quad%
    \begin{subfigure}[b]{0.32\textwidth}
          \centering
         \includegraphics[width=\textwidth]{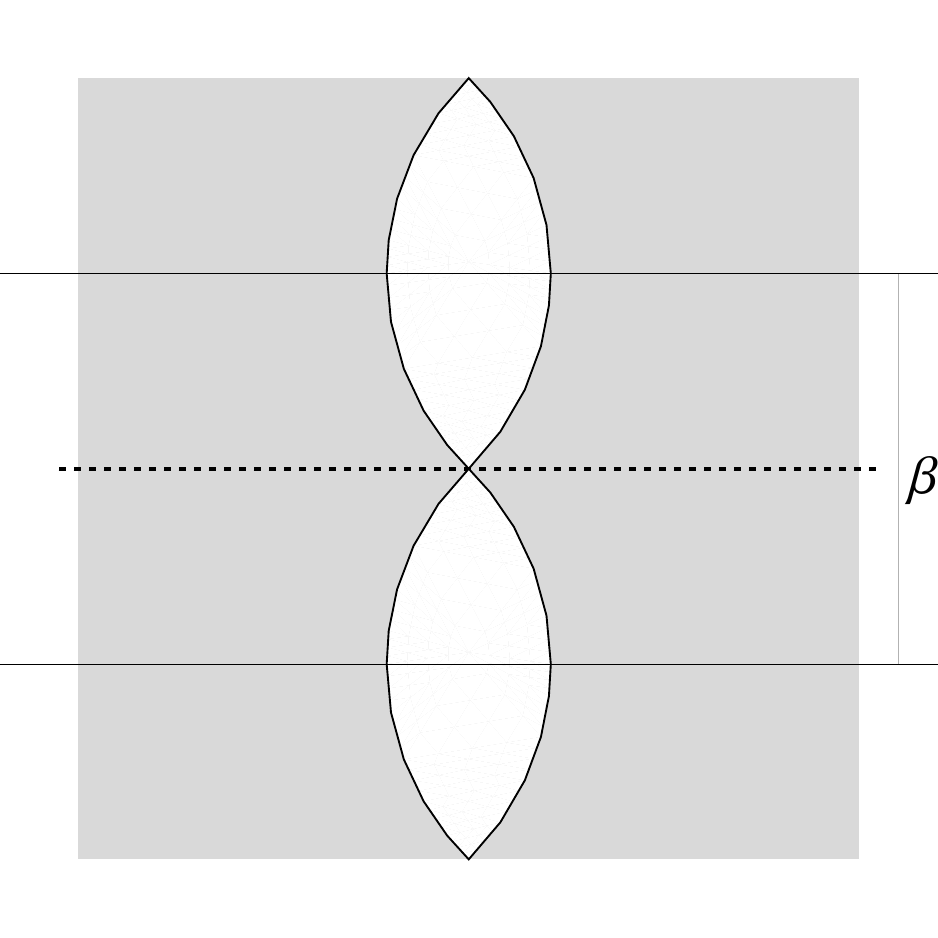}
         \subcaption{$T\gtrsim T_0$}
    \end{subfigure}
    \hfill
    \begin{subfigure}[b]{0.32\textwidth}
          \centering
         \includegraphics[width=\textwidth]{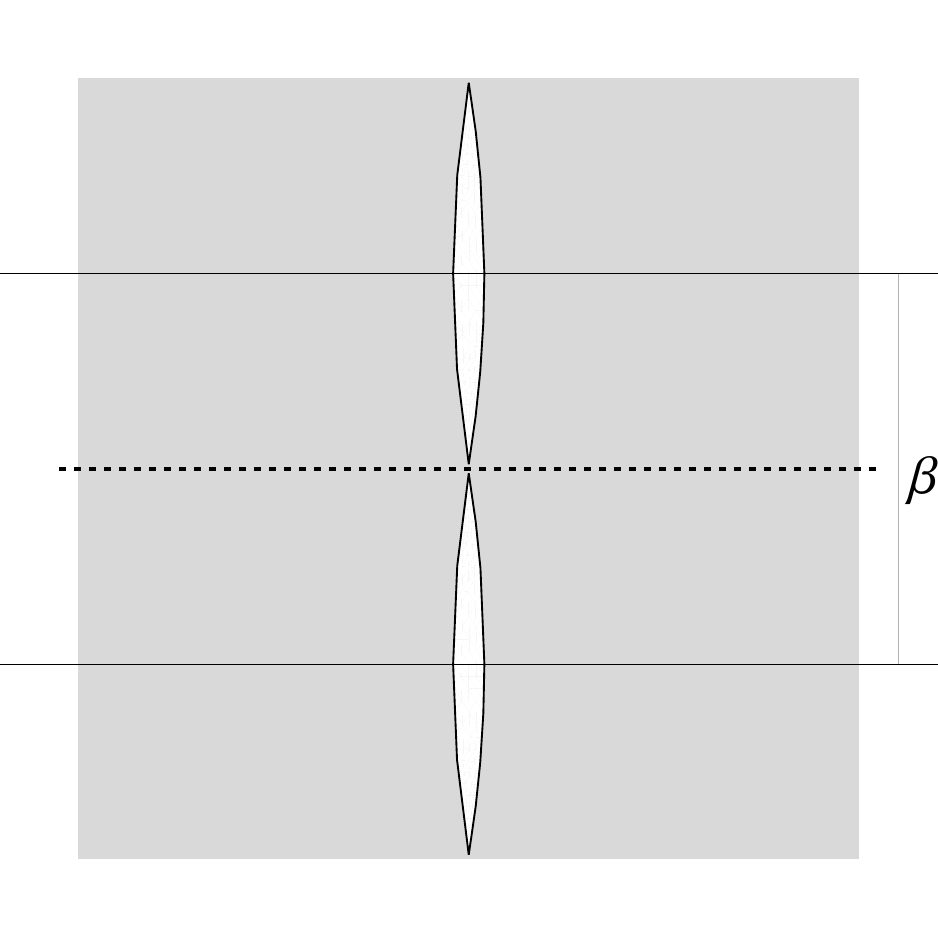}
         \subcaption{$T \gg T_0$}
    \end{subfigure}
    \hfill
    \caption{Cartoon of the instanton solutions (white) on the Euclidean worldsheet (grey) with period $\beta = 1/T$, in different temperature regimes, relative to $T_0=\mu/(2m_M)$, which is the inverse size of the vacuum instanton.}
    \label{fig:instantons}
\end{figure}

As the temperature approaches $T_0$, the instanton becomes sensitive to the periodicity of the Euclidean worldsheet and is distorted. Qualitatively, the shape of the instanton solution changes from a circle ($T < T_0$) to two arc segments ($T \gtrsim T_0$), as illustrated in Figure~\ref{fig:instantons}. This deformation modifies both the prefactor and the bounce action. 

The temperature dependence of the prefactor $C(T)$ is complicated and was analysed in detail in \cite{Ivlev87}. For our purposes, it suffices to note that  between $T_0$ and $T_\star \sim \sqrt{\mu}$ it increases only by a factor of $\mathcal{O}(1)$. More generally, in $(1+1)$ dimensions, the prefactor is never exponentially enhanced relative to its zero-temperature value \cite{Garriga_1994}; hence, the only significant increase in the tunnelling rate comes from the bounce action. 

In $(1+1)$ dimensions, the bounce action can be written as
\begin{equation}
    S_B  = 2 m_M L - \mu A \, ,
\end{equation}
where $L$ is the length of the worldline of a monopole (or antimonopole), and $A$ is the area of the enclosed region of true vacuum. The relevant instanton solution when $T > T_0$ is shown in Figure~\ref{fig:instantons}(b) \footnote{This solution assumes point-like monopoles; introducing the monopole size $\delta$ smooths out the kinks where the two arches meet but does not change the bounce action provided $\delta \ll r_M$ (or equivalently $\kappa \gg 1$). For a detailed discussion, see Ref.~\cite{Ivlev87}.}. Following \cite{Garriga_1994}, we interpret this solution as representing a combination of quantum tunnelling through a potential barrier and thermal hopping towards its top. Subcritical bubbles of true vacuum are constantly nucleated over the worldsheet due to thermal fluctuations; these bubbles can then tunnel to supercritical size through quantum fluctuations, and expand to infinity. The instanton consists of two arc segments of radius $r_M$ and length $L = \alpha \, r_M$, where 
\beq
\alpha = 2 \arcsin{(\beta/2r_M)}\,,
\eeq
enclosing an area 
\beq
A = r_M^2 (\alpha - \sin{\alpha})\,.
\eeq
Therefore, the bounce action is given by
\begin{equation}\label{eq:SBT}
    S_B = \frac{2m_M^2}{\mu} \left[ \arcsin{\left(\frac{T_0}{T}\right)} + \frac{T_0}{T} \sqrt{1 - \left( \frac{T_0}{T}\right)^2} \right]\, .
\end{equation}
For typical values $m_M^2/T^2\sim m_M^2/\mu\gtrsim 10$ of phenomenological interest, the finite-temperature breaking rate is dramatically faster than the zero-temperature breaking rate of Eq.~\eqref{eq:4} since the action scales as $m_M/\sqrt{\mu}$ rather than $m_M^2/\mu$.

Equation~\eqref{eq:SBT} holds in the thin-string limit. When $T \gtrsim T_{\rm  ts} \equiv \sqrt{\mu}/\kappa^{1/4}$, the width of the instanton becomes comparable to the thickness of the string $\sim 1/\sqrt{\mu}$, and the thin-string approximation breaks down. This regime is relevant for the phenomenology of the string network since typically $T_\star >  T_{\rm  ts}$. While the exact functional form of $S_B(T)$ may be different in this case, we expect that $S_B(T \gg T_0) \sim 2m_M/T$ and ${\rm d}S_B(T)/{\rm d}T \leq0$, such that Equation~\eqref{eq:SBT} should still give a reasonable estimate up to $\mathcal{O}(1)$ differences. Nevertheless, in our analysis we will also consider a more conservative estimate for the decay rate, assuming Eq.~\eqref{eq:SBT} only for $T < T_{\rm ts}$ and no decay otherwise. As we will see, this choice does not affect our qualitative conclusions. A detailed calculation of the decay rate at finite $T$ and with finite string thickness would be very helpful in reducing the uncertainties on this point, and is left for future work.

For completeness, let us remark that once the finite wall thickness $\delta$ is taken into account, at all $T$ there exists another solution alongside the periodic instanton, consisting of two parallel worldlines separated by a distance $\sim \delta$. This ``static" solution is associated to a sphaleron process (pure thermal hopping), and gives a contribution to the decay rate of $\Gamma_{\rm sp} \sim \exp(-F_{\rm sp}/T)$, where $F_{\rm sp} \approx 2m_M$ is the free energy of the sphaleron configuration (neglecting the wall thickness misses this sphaleron, since $F_{\rm sp}$ would have no extremum) \cite{Garriga_1994}. Importantly, we note that the rate $\Gamma_{\rm sp}$ is always subdominant compared to the rate given by Eq.~\eqref{eq:SBT}. In the high temperature limit the decay rates given by the periodic instanton and the sphaleron coincide ($S_B \sim 2m_M/T$ when $T \gg T_0$), and indeed the two configurations tend to the same form. Specifically, once the finite wall thickness becomes comparable to the radius of the instanton at $T\sim \sqrt{\mu}$, the cusps in Figure~\ref{fig:instantons}(c) are smoothed out, and this solution reduces to the static, parallel sided one.

These properties are specific to false vacuum decay in $(1+1)$ dimensions. Instead, for $(1+d)$ dimensions with $d > 1$, the sphaleron exists even in the thin wall limit, and becomes the dominant contribution above some finite $T$. Additionally, at sufficiently high temperatures the instanton ceases to exist.


In the discussion above we have neglected the possible temperature dependence of the VEV $v$. For a typical second order phase transition, we expect that as $T \to T_\star$, $v(T) \to 0$ such that $S_B \to \infty$ and the network cannot decay immediately after formation. Since we wish to remain agnostic about the details of the phase transition, we do not include this effect in our estimates below. Nevertheless, we expect it to only mildly affect our quantitative conclusions in the regime under consideration, as discussed in Appendix~\ref{app:tunnelling}.

\subsection{Impact of finite-temperature breaking on the network}\label{subsec:finiteT}

The enhancement of the decay rate at $T\gtrsim T_0$ significantly affects the cosmological evolution of a metastable string network produced by a finite-temperature phase transition. 

The network forms at temperature $T_\star \simeq v$ via the percolation of proto-string segments of length $\xi$. For $\kappa \gg 1$ (the regime relevant for metastability), $T_\star > T_0$, which means that the decay rate at formation is exponentially larger than its zero-temperature value. Assuming that the Universe is radiation dominated, for $T>T_0$, the number of decays per Hubble time per Hubble length of string including temperature corrections, $N_H(T)$, is given by
\begin{equation}\label{eq:networkrateT}
    N_H(T) = \Gamma(T) H^{-2} = \frac{8 \kappa^2} {\pi} \frac{1}{G_{\rm N} \mu} \left( \frac{T_0}{T}\right)^4 \exp{\left[ - S_B \left( \kappa, \frac{T_0}{T} \right)  \right]} \, .
\end{equation}

Immediately after the network forms, each correlation volume $\xi^3$ contains a string length of order $\xi$. On a timescale of order the Hubble time, the network then approaches a scaling solution, with roughly one Hubble length of long strings per Hubble volume, by converting long-string length to small loops that quickly decay. 
During this period, strings break at the rate given by Eq.~\eqref{eq:networkrateT}. Some of the resulting monopole-antimonopole pairs lie on string regions that, in the absence of breaking, would have been radiated as sub-horizon loops; in such cases the resulting segments are sub-horizon and quickly decay with the monopole-antimonopole pairs annihilating. However, other monopole-antimonopole pairs lie on the would-be remaining infinite strings, which therefore become finite-length.

If $N_H$ is order one, every string segment (including those that would otherwise be infinite) quickly breaks into sub-horizon pieces, and the entire network decays within a single Hubble time.

We are instead interested in the limit $N_H  \ll 1$, corresponding to $S_B(\kappa,T_\star)\gg 1$.  In this regime, because $S_B$ increases rapidly as the temperature falls, the probability $p$ that a given Hubble length of a long string breaks is well approximated by the breaking that occurs in the first Hubble time after formation: $p\simeq N_H(T_\star)$. In the absence of small couplings in the underlying theory, the formation temperature satisfies $T_\star\simeq \mu^{1/2}$. The resulting segments are super-horizon but finite, with typical length
\beq \label{eq:elli}
\ell_i \simeq \frac{1}{p} H(T_\star)^{-1} =  H(T_\star)^{-1} N_H(T_\star)^{-1}~. 
\eeq
$N_H$ may equivalently be interpreted as the probability of finding a monopole-antimonopole pair on which two long string segments end in a particular Hubble patch soon after network formation. 

Crucially, the breaking events on a would-be infinite string are independent, occurring as a Poisson process per unit length. Therefore, the lengths of the resulting segments follow a distribution $P(\ell)\sim e^{-\ell/\ell_i}$ with strings of length $\ell\gg \ell_i$ exponentially rare.  As a result, soon after formation the network contains no infinitely long strings.

For definiteness, we take $\mu = 2\pi v^2$ and $T_\star= v$, as in a critical Abelian-Higgs string with symmetry breaking scale $v$.  Eq.~\eqref{eq:elli} then simplifies to
\beq
\ell_i \simeq \frac{1}{M_{\rm Pl}} \exp \left( \sqrt{8\pi \kappa}\right)~,
\eeq
where we drop order-one numbers from the prefactor. We cannot determine $\ell_i$ more precisely than this, since some segments might recombine to form longer strings (with short pieces radiated away). Moreover, the detailed dynamics of percolation, loop production, and loop decay introduce additional order-one uncertainties. However, our conclusions are driven by the form of the exponential, which is robust.

One subtlety is that the quantity most relevant for the subsequent evolution of a long string segment is not its total initial length $\ell_i$, but the typical separation between its endpoints  $\ell_0$. For example, $\ell_0$ sets the time when the two endpoints enter each other's horizon, at which point the segment decays. At formation, the strings follow random-walk trajectories, so $\ell_0$ is parametrically smaller than $\ell_i$.  Approximating that, soon after formation,  the strings are roughly straight on scales of order the Hubble length $H_\star\equiv H(T_\star)$, we obtain
\begin{equation}\label{eq:l0finiteT}
    \ell_0 \simeq \left( \frac{\ell_i}{H_\star} \right)^{1/2} \simeq \frac{1}{v} e^{\sqrt{2\pi\kappa}} \, ,
\end{equation}
where, as before, our main interest is the exponential dependence rather than the prefactor. 

It is useful to briefly compare this to what would happen if the strings broke only at the $T=0$ rate. In that case, some breaking would still occur during the first Hubble time after network formation, and, by the same argument as above, there would be no infinite strings. However, without the temperature dependence, the rate of breaking $\Gamma$ remains constant while the Hubble time and Hubble length grow. As a result, the network is ultimately destroyed when $\Gamma H^{-2}\sim 1$, i.e., when each Hubble length of string breaks within a Hubble time, with the breaking during the final Hubble time of the network's life being the most important. 

The situation changes once finite-temperature effects are included, since, as discussed, the bounce action at formation
\beq \label{eq:compareSB}
\begin{aligned}
S_B(T_\star)& \simeq \sqrt{8\pi\kappa} ~,
\end{aligned}
\eeq
is much smaller than the zero-temperature value
\beq \label{eq:compareSB2}
S_B(T=0)= \pi \kappa~,
\eeq
in the metastable-string regime $\kappa\gtrsim 10$. Whether this initial breaking dominates over the late-time, $T\simeq 0$, breaking  (which proceeds more slowly but over a much longer duration) is a numerical question. In Section~\ref{sec:late} we will see that it does, which is plausible given that we are comparing an exponential enhancement with a power-law increase due to the increased Hubble time at late times.

\begin{figure}[t]
    \centering
    \includegraphics[width=0.7\textwidth]{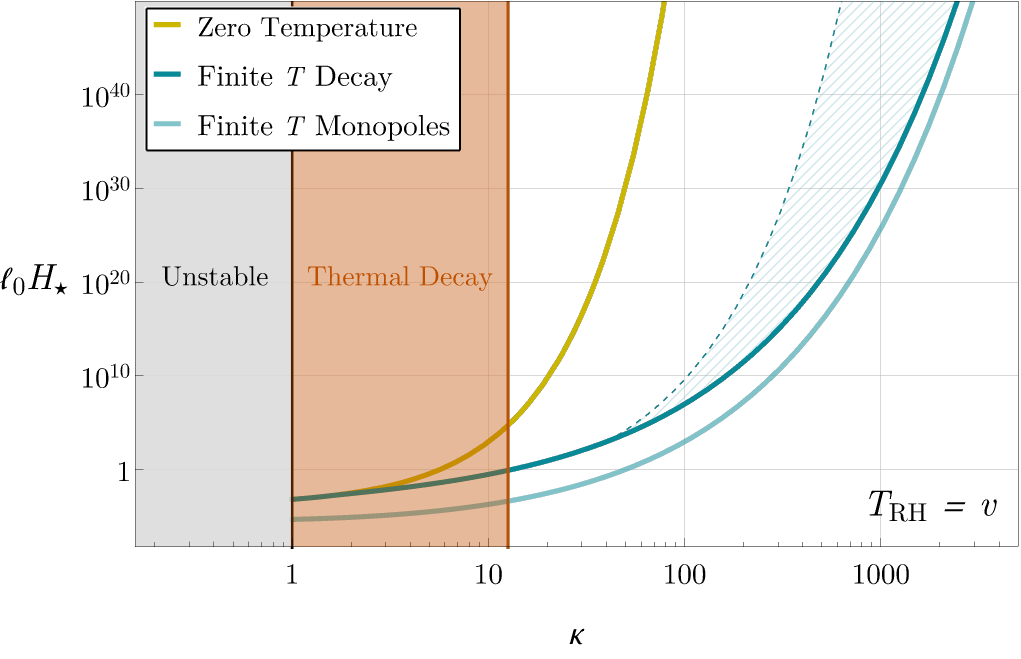}
    \caption{The typical distance between string endpoints, $\ell_0$, soon after network formation, in units of the Hubble length at that time $H_\star^{-1}$. The results are plotted as a function of $\kappa\equiv m_M^2/\mu$, where $m_M$ is the monopole mass and the string tension $\mu$ is fixed to satisfy $G_{\rm N}\mu = 10^{-7}$. The yellow line shows the endpoint separation if breaking occurred only at the zero-temperature rate ($T = 0$). The blue lines include finite-temperature effects, due to the enhanced breaking rate (Eq.~\eqref{eq:l0finiteT}) and pre-existing monopoles produced from the thermal bath (Eq.~\eqref{eq:ell0}). For the latter, we conservatively assume the correlation length $\xi \simeq T_\star^{-1}$ and $T_{\rm RH} \simeq v$. The dashed line shows the upper bound on $\ell_0$ obtained assuming the tunnelling action from Eq.~\eqref{eq:nothinstring}, in units of Hubble when $T = T_{\rm ts}$. The red shaded region indicates values of $\kappa$ for which typical segments lie within a single Hubble patch and the network decays rapidly.
    }
    \label{fig:l0_thermal}
\end{figure}

Figure~\ref{fig:l0_thermal} shows the initial endpoint separation of long segments, $\ell_0$,  as a function of $\kappa$ due to thermal breaking. For comparison, we also plot what the segment length would be one Hubble time after formation if breaking occurred only at the $T=0$ rate. We also plot the length of segments due to pre-existing monopoles, which we analyse in the next subsection. For $\kappa\simeq 60$, as typically assumed in metastable-string studies, we find $\ell_0\lesssim 10^4\,H_\star^{-1}$, implying that the segments are super-horizon but only moderately so.

As previously mentioned, our expression for $S_B(T)$ is not fully reliable when $T > T_{\rm ts}$ and the thin-string approximation breaks down. Assuming conservatively that the strings only start decaying when $T = T_{\rm ts}$, the analysis above remains unchanged except for the bounce action which becomes
\begin{equation}\label{eq:nothinstring}
    S_{B}(T_{\rm ts}) \simeq 2 \kappa^{3/4} \, ,
\end{equation}
which  for large $\kappa$ is smaller than the zero-temperature value. The resulting endpoint separation is also shown in Figure~\ref{fig:l0_thermal}, with the shaded region corresponding to the uncertainty arising from the value of $S_B(T)$ outside the thin-string regime.

\subsection{Monopoles produced by freeze-in} \label{ss:thermalM} 

The lengths of the long string segments are further reduced if monopoles are already present when the network percolates. Any random-walk trajectory of a proto-string that encounters a monopole terminates, halting its growth.  We assume that the monopole population is produced thermally after reheating at temperature $T_{\rm RH}$, with any monopoles from earlier epochs diluted by inflation and negligible. Specifically, we require $ T_{\rm RH}\lesssim m_M$ to avoid a monopole-producing symmetry-breaking transition (while still imposing $v\lesssim T_{\rm RH}$ so that strings form); otherwise the monopole number density would be large, segments would be at most a few Hubble lengths, and the network would decay rapidly.  We also assume that the correlation length at the time of the transition is not much larger than the typical distance between the thermally produced monopoles (which is no larger than the Hubble distance). When this is not the case, it might be that all monopole-antimonopole pairs would be connected by short strings and quickly decay away.

With such a reheating temperature,  monopole production is Boltzmann suppressed with rate per unit volume $dn_M/dt \sim e^{- E_{\rm tot}/ T} $, where \(E_{\rm tot}\) is the total energy of the monopole pair produced. This includes their mass, kinetic energy, and binding energy from long-range magnetic interaction between them. While the kinetic energy leads to a suppression of the production rate, the binding energy enhances it. For monopoles which are sufficiently separated or created with enough kinetic energy to survive for a significant period of time after production, the binding energy is smaller than the kinetic energy. However, due to the suppression of the production rate of monopoles with excessive kinetic energy
, the majority of monopoles of interest will have \(E_{\rm tot}\) close to the sum of the masses of the two monopoles. This motivates a simplification for the exponent of the estimated number density of frozen-in monopoles, on which the (long-lived) cosmic strings can end, such that \(E_{\rm tot} \approx 2 m_M \).

The prefactor of the exponent depends on the thermal average $\left<\sigma v \right>$ of the monopole-antimonopole production cross section. Following \cite{Turner1982}, we parametrise
\begin{align} \label{eq:GammaV}
    \frac{dn_M}{dt} 
    \simeq a h^2 \left( \frac{m_M}{T} \right)^p m_M^3 T e^{-2 m_M / T}\,,
\end{align}
where $a$ and $p$ depend on the model and the environment, and $h$ is the monopole magnetic charge (quantized in units of $2\pi/e$ with $e$ the minimal electric charge). We neglect an additional contribution to the prefactor counting the number of production/annihilation channels. 

Two limiting forms of the prefactor in Eq.~\eqref{eq:GammaV} arise depending on the monopole mean free path $\lambda$  relative to the capture distance $a_c = h^2/(4 \pi T)$ \cite{Preskill1979}.  The mean free path depends on the charged-particle plasma. If the plasma contains many relativistic degrees of freedom, $N_q$, so that scattering is frequent (for example, at  $T\simeq m_M/60$ when $h^4 N_q \gtrsim 2 \times 10^5$), one finds $p = 0$. If the mean free path is long then $p = -11/10$. Details can be found in  Refs.~\cite{Turner1982,Preskill1979,Collins1984}.

Assuming instantaneous reheating to $T_{\rm RH}$, a reasonable estimate of the monopole number density at string formation is obtained by integrating the production rate over one Hubble time starting from $T_{\rm RH}$. 
For definiteness, we take $p=0$ and $a h^2\sim \mathcal{O}(1)$; since these choices affect only the prefactor, they do not make a major difference to our subsequent results. 
The monopole number density one Hubble time after reheating is then approximately
\beq
n_M \simeq \frac{m_M^3 M_{\rm Pl}}{T_{\rm RH}} e^{-2m_M/T_{\rm RH}}~,
\eeq
and including redshifting to the temperature of string formation gives
\beq \label{eq:NmTv}
n_M(T=v) \simeq \frac{m_M^3 v^3 M_{\rm Pl}}{T_{\rm RH}^4} e^{-2m_M/T_{\rm RH}}  ~.
\eeq

In more realistic cosmological histories, the radiation bath reaches a maximum temperature $T_{\rm max}> T_{\rm RH}$ before reheating completes, even in the case of perturbative reheating \cite{Giudice:2000ex}, and non-perturbative effects can further complicate the dynamics (see \cite{Amin:2014eta} for a review). In this case, provided $T_{\rm max}$ is below the monopole symmetry breaking scale, our main conclusions continue to hold with $T_{\rm RH}$ replaced by $T_{\rm max}$, although the prefactors are modified, typically by powers of $T_{\rm RH}/T_{\rm max}$. 
Additional effects, such as inhomogeneities during reheating, non-perturbative preheating, oscillations of the monopole mass while the symmetry-breaking scalar has not settled into its minimum  \cite{Collins1984,Chung:1998rq} could also significantly enhance monopole production leading to shorter string segments. Similarly, it might be that gravitational production of monopoles due to the oscillating background, analogous to the fermion production analysed in \cite{Ema:2018ucl}, could be efficient if the inflaton mass is larger than the monopole mass; we will study this process in future work.

\subsubsection{Percolation}\label{sss:perc}

Given a monopole abundance $n_M$ at string formation, we can estimate the lengths of the long-string segments as follows. At the phase transition, the symmetry-breaking field (e.g. a complex scalar) is uncorrelated on scales larger than the correlation length $\xi$. On general grounds $T_\star^{-1}\lesssim \xi\lesssim H_\star^{-1}$, with the precise value depending on the details of the crossover or phase transition (see, e.g., \cite{Rajantie:2001ps}). As discussed, the percolation of strings can be viewed as a random walk in which segments of typical length $\xi$ (one per $\xi^{3}$ volume) join together. If the random walk enters a volume $\xi^{3}$ containing a monopole or antimonopole, it has an order-one probability of terminating there, since the monopole necessarily has an attached string segment carrying magnetic flux.

To a first approximation, we treat the monopole and antimonopole positions as random and uncorrelated. 
Consequently, the probability that a string segment encounters a monopole within a given $\xi^3$ volume is
\beq
p_{M} \simeq \xi^3 n_M~.
\eeq
Hence the typical segment length at this stage (before finite-temperature breaking) is
\beq \label{eq:elli2}
\ell_i \simeq \xi^{-2}n_M^{-1}~. 
\eeq
Since encounters in successive $\xi^3$ volumes are independent, segments much longer than $\ell_i$ are exponentially rare, and there are already no infinitely long strings at this time.  
As in the case of finite-temperature breaking, the random walk structure of the initial configuration implies that the typical endpoint separation $\ell_0$ is smaller than $\ell_i$ in Eq.~\eqref{eq:elli2}, with
\beq
\ell_0 \simeq \xi \left(\xi^{3}n_M\right)^{-1/2}~.
\eeq

There will also be closed string loops, typically shorter than $\ell_i$, arising from random walks that return to their starting point before encountering a monopole. As with open strings, closed loops with lengths greater than $\ell_i$ are exponentially rare. 
Results from random-walk theory imply that when the monopole number density is low, roughly one-third of strings are closed loops, while at higher densities this fraction decreases (see also Section~\ref{sec:inflation}, where we find a loop fraction of order $1 \%$ for $\xi^3 n_M = 0.1$). The typical radius of these loops is smaller than $\ell_0$.

We assume Eq.~\eqref{eq:NmTv} for the initial monopole abundance. The endpoint separation $\ell_0$ is maximised when $T_{\rm RH}\simeq T_\star\simeq v$, so
\beq \label{eq:l0TM}
\ell_0 \lesssim  \left( \frac{v}{m_M^3 M_{\rm Pl} \xi}\right)^{1/2} e^{m_M/v} \simeq (2 \pi \kappa)^{-3/4} \frac{1}{\sqrt{M_{\rm Pl} \xi}} \frac{1}{v}  e^{\sqrt{2\pi \kappa}} ~,
\eeq
where the expression on the right applies to a critical Abelian-Higgs string. To find the maximum possible endpoint separation, we take $\xi = T_\star^{-1} \simeq v^{-1}$, its minimal allowed value, which gives
\beq \label{eq:ell0}
\ell_0 \lesssim (2 \pi \kappa)^{-3/4} \frac{1}{\sqrt{M_{\rm Pl} v}} e^{\sqrt{2\pi \kappa}}~.
\eeq
For $\kappa = 64$ and $G_{\rm N}\mu = 10^{-7}$, Eq.~\eqref{eq:ell0} leads to $\ell_0 \lesssim 30H_\star^{-1}$, so the resulting segments are at most only marginally super-horizon. Comparing with Eq.~\eqref{eq:l0finiteT}, when $T_{\rm RH}\simeq v$ the pre-existing monopoles dominate over finite-temperature breaking because of the smaller prefactor, as shown in Figure~\ref{fig:l0_thermal}. For $T_{\rm RH}\gg v$, pre-existing monopoles become exponentially more important, since the exponential in Eq.~\eqref{eq:l0TM} contains $m_M/T_{\rm RH}$ rather than $m_M/v$.

Our assumption that the monopole and antimonopole locations are uncorrelated is only approximate, since they are produced in pairs and are therefore typically close together. However, provided their separation exceeds a correlation length, any given pair has a less than order-unity probability of joining to form a short segment. This effect therefore only influences the already-uncertain prefactors, without altering the exponential dependence. We demonstrate this numerically in Section~\ref{sec:inflation} in the context of monopoles produced during inflation, by studying random walks with a realistic monopole distribution.

We conclude that pre-existing monopoles are likely to dominate over finite-temperature breaking, although the former involves greater model-dependence and uncertainty in the prefactors. For reheating temperatures not far above the symmetry breaking scale, the key exponential is similar for both mechanisms, since in each case it reflects the Boltzmann suppression for producing a monopole-antimonopole pair.

\section{Networks formed by inflationary fluctuations} \label{sec:inflation}

During inflation with Hubble scale $H_I$, any field lighter than $\sqrt{2}H_I$ has sub-horizon fluctuations of amplitude $H_I/(2\pi)$. 
As these fluctuations leave the horizon, the coarse-grained field (on scales of order $H_I^{-1}$) undergoes a random walk with a step size of $H_I/(2\pi)$ per e-fold. 
If the string symmetry breaking scale satisfies $v\lesssim H_I/(2\pi)$ and the radial mode is light compared to $H_I$, the symmetry-breaking field is therefore randomised during inflation. Specifically, at the end of inflation the field is random on spatial scales larger than $H_I^{-1}$. As the radial mode later relaxes to its vacuum expectation value, non-trivial winding around the vacuum manifold occurs over the same length scale. Such winding produces proto-string segments with a typical length of order $H_I^{-1}$ and number density of order $H_I^{3}$, which connect and percolate.

If $H_I/(2\pi)$ also exceeds the scale of the first symmetry-breaking, there are also field configurations that relax to monopoles, with a comparable number density of order one per $H_I^{-3}$ volume. In this scenario a network of long strings never forms, because the string segments often encounter a monopole and terminate without percolating.

We instead assume that $H_I/(2\pi)$ lies between the two breaking scales, so that many proto-string segments form but the fields involved in the first breaking remain in their vacuum. We also assume that the post-inflationary temperature remains well below both breaking scales, otherwise the finite-temperature effects discussed in Section~\ref{sec:reheat} would be relevant.

\subsection{Monopole abundance from inflation} 

Even though the symmetry associated with the monopoles is not restored during inflation in the regime we consider, monopole-antimonopole pairs are occasionally produced by exponentially rare large fluctuations. This production can be viewed as being due to thermal fluctuations at the de Sitter horizon, which has temperature $T_{\rm dS} = H_I/2\pi$; the rate of sufficiently large fluctuations is therefore Boltzmann suppressed and proportional to $\exp\left(-m_M/T_{\rm dS} \right)$.\footnote{More precisely, this picture is valid as long as the dynamics occur within a static patch. In our case, the monopoles are small compared to the Hubble parameter during inflation, so this is consistent.} 
The production rate per unit volume and time, $\Gamma$, and the resulting monopole number density at the end of inflation can be computed following Ref.~\cite{Basu:1991ig}.

The calculation is similar to that for monopole-antimonopole nucleation on the string worldsheet discussed in Section \ref{sec:reheat}. In the present case, the relevant instanton is a circle of radius $r_0 = H_I^{-1}$, embedded in a 4-sphere, and the tunnelling action is $S_B = 2\pi m_M H_I^{-1}$. The result for $\Gamma$ is approximately given by
\begin{equation} \label{eq:Gammainf}
    \Gamma \simeq H_I^4 e^{-2\pi m_M/H_I} \, .
\end{equation}
The number of monopole-antimonopole pairs nucleated during an interval ${\rm d}t$ within a physical volume $R(t)^3 {\rm d}x^3$ is then given by
\begin{equation}
    {\rm d}N_M = \Gamma e^{3H_I t} {\rm d}^3 x {\rm d}t \, ,
\end{equation}
where the scale factor during inflation is $R(t) = e^{H_I t}$. The physical separation $d$ between a particular monopole-antimonopole pair at a later time $t'$ is given by~\footnote{Note that within the semiclassical treatment described here there is no well-defined notion of a nucleation time; equations like Eq.~\eqref{eq:mmdistance} seem to suggest that monopole-antimonopole pairs are nucleated only when $t \rightarrow -\infty$. This signifies a breakdown of the classical treatment very close to the moment of nucleation. However, we are interested in the monopole distribution at the end of inflation, long after the nucleation event for most monopoles, when the semiclassical treatment is reliable. This point is discussed further in Ref.~\cite{Basu:1991ig}. }
\begin{equation}\label{eq:mmdistance}
    d = 2r = 2 H_I^{-1} \sqrt{e^{2H_I (t' - t)} +1} \, .
\end{equation}
Using these results, we can derive the density distribution of monopoles as a function of the monopole-antimonopole separation $r$
\begin{equation}\label{eq:Rdist}
    \frac{{\rm d} n_M}{ {\rm d} r} =  e^{-3H_I t} \frac{{\rm d} N_M}{ {\rm d} r \, {\rm d}^3x} = \frac{H_I^5 r}{(H_I^2 r^2 - 1)^{5/2}} e^{-2\pi m_M/H_I} \, .
\end{equation}
The distribution diverges at $r = H_I^{-1}$, signifying a breakdown of the semiclassical approximation close to the time of nucleation, which is a strictly quantum-mechanical process. This introduces a cutoff in the distribution, $r_{\rm min} = H_I^{-1} + \Delta$, since pairs closer than twice this radius have not yet nucleated by the end of inflation. To estimate $\Delta$ we adopt the same approach as Ref.~\cite{Basu:1991ig}: we analytically continue the tunnelling action to $r = H_I^{-1} + \Delta$, interpret the resulting expression as a quantum-mechanical phase and find the value of $\Delta$ for which this phase becomes large. We obtain $\Delta \lesssim H_I^{-1}$ provided $H_I < m_M$, with typical values $\Delta \gtrsim 10^{-2} H_I^{-1}$. The exact value of the cutoff is unimportant for our purposes, as it only affects the prefactors in our final expressions by an $\mathcal{O}(1)$ number. 

Then, integrating Eq.~\eqref{eq:Rdist} from $r_{\rm min}$ to infinity, we obtain the monopole abundance at the end of inflation, $n_M$,
\begin{equation}\label{eq:nMHI}
    n_M \simeq H_I^3 e^{-2\pi m_M/H_I} \, .
\end{equation}
Given our assumed range for $H_I$, Eq.~\eqref{eq:nMHI} provides a lower bound on $n_M$ of
\begin{equation} \label{eq:nMlower}
    H_I \geq (2\pi) v  \implies n_M \geq (2\pi v)^3 e^{- m_M/v} \, .
\end{equation}
For the particular case of the critical Abelian-Higgs string, with $\mu=2\pi v^2$, this  becomes
\begin{equation} \label{eq:lowernM}
    n_M \geq (2\pi \mu)^{3/2} e^{- \sqrt{2\pi \kappa}} \, .
\end{equation}
This result has an exponential dependence on $\sqrt{\kappa}$ (similar to the thermal nucleation case), in contrast to the vacuum string-decay rate, which depends exponentially on $\kappa$, see Eq.~\eqref{eq:4}.

\subsection{Percolation of the string network} \label{sec:perc3}

Similarly to Section~\ref{sss:perc}, the symmetry-breaking field is uncorrelated on scales larger than $H_I^{-1}$, and string percolation can be viewed as a random walk with step size of order $H_I^{-1}$, joining segments that arise roughly once per $H_I^{-3}$ volume. The  presence of even exponentially rare monopoles terminates these random walks, preventing the formation of infinitely long strings.

As a first approximation, we again treat the locations of the monopoles and antimonopoles as random and uncorrelated. More precisely, the physical separation between a monopole and its partner goes as $2H_I^{-1} e^{\Delta N}$, where $\Delta N$ is the number of e-folds between nucleation and the end of inflation. Pairs nucleated during the last e-fold of inflation remain close together and often connect to each other by a short string segment, whereas pairs nucleated earlier are separated by a distance of at least several $H_I^{-1}$, over which the symmetry-breaking field is completely randomised. Consequently, the random walk beginning on a monopole fails to find its corresponding antimonopole at least an order-one fraction of the time. Neglecting the nearby pairs changes the prefactor in Eq.~\eqref{eq:nMHI} by an order-one amount, leaving the crucial exponential dependence on $m_M/H_I$ unchanged. 

Since in this section we are considering reheating temperatures below $v$, it is reasonable to assume that the Universe has an era of matter domination after inflation, as the inflaton oscillates in a quadratic potential before decaying. The string network percolates once the Hubble parameter falls to $H\simeq v$, at which point the radial mode relaxes to its vacuum expectation value. The monopole number density given by Eq.~\eqref{eq:nMHI} is therefore reduced by a redshift factor $(H_I/v)^3$, but this only affects the prefactor.   
We then simply repeat the analysis of  Section~\ref{sss:perc}, now taking the correlation length $\xi$ to be $H_I^{-1}$. The resulting typical long-string length is
\beq
\ell_i \simeq H_I^{2}n_M^{-1} \simeq \frac{H_I}{v^2} e^{2\pi m_M/H_I}  ~, 
\eeq
and longer segments are exponentially rare. 
The initial length distribution can be obtained from the expected number of steps $N$ of a random walk that has a probability $p_M = \xi^3 n_M$ of terminating at each step, with the walk length $\ell = N\xi$. The normalised distribution is therefore
\begin{equation}\label{eq:segmentdistribution}
    P(\ell) = \frac{\ell}{\ell_i^2}\,e^{-\ell/\ell_i}\, .
\end{equation}
As in Section~\ref{sss:perc}, due to the random walk, the typical direct distance between the endpoints is
\beq
\ell_0 \simeq v^{-1}e^{\pi m_M/H_I}~.
\eeq
For a critical Abelian-Higgs string, taking the minimal allowed value $H_I=2\pi v$ leads to
\beq
\ell_0 \lesssim v^{-1} \, e^{\sqrt{\pi\kappa/2}} ~.
\eeq
Not surprisingly, the exponential dependence on $\sqrt{\kappa}$ is similar to that found for monopole production from the thermal bath, Eq.~\eqref{eq:ell0}. 

\subsection{Numerical results from random walk statistics}

The conclusions of Section~\ref{sec:perc3} can be validated numerically by modelling the percolation of the string network as a self-interacting random walk. In the spirit of Ref.~\cite{COPELAND1988445}, we do so by implementing an improved Vachaspati-Vilenkin algorithm \cite{VachaspatiVilenkin,Strobl97}, which simulates the formation of topological defects using a Monte Carlo approach. The algorithm assigns a random field value (taken from the vacuum manifold of the broken phase) to each point on a lattice, with the lattice spacing set by the correlation length $\xi$. This models the fact that field values are uncorrelated over distances of order $\xi$. Each plaquette with non-trivial winding is taken to contain a string segment, which necessarily joins with other segments in nearby plaquettes to form either an infinite string or a loop. These simulations allow us to extract statistical properties of the string network soon after formation. 

More specifically, our analysis is reminiscent of  Ref.~\cite{COPELAND1988445}, which studied string networks formed alongside a simultaneous monopole-producing  phase transition, leading to a large monopole number density. However, unlike that work, we sample the spatially discretised  field only from the $\mathrm{U}(1)$ vacuum manifold, so monopoles do not automatically form and must be introduced on the lattice ``by hand". This approach allows us to more easily control the monopole number density $n_M$, and access the low densities relevant for our work. 
\begin{figure}[t]
    \centering
    \includegraphics[width=0.5\textwidth]{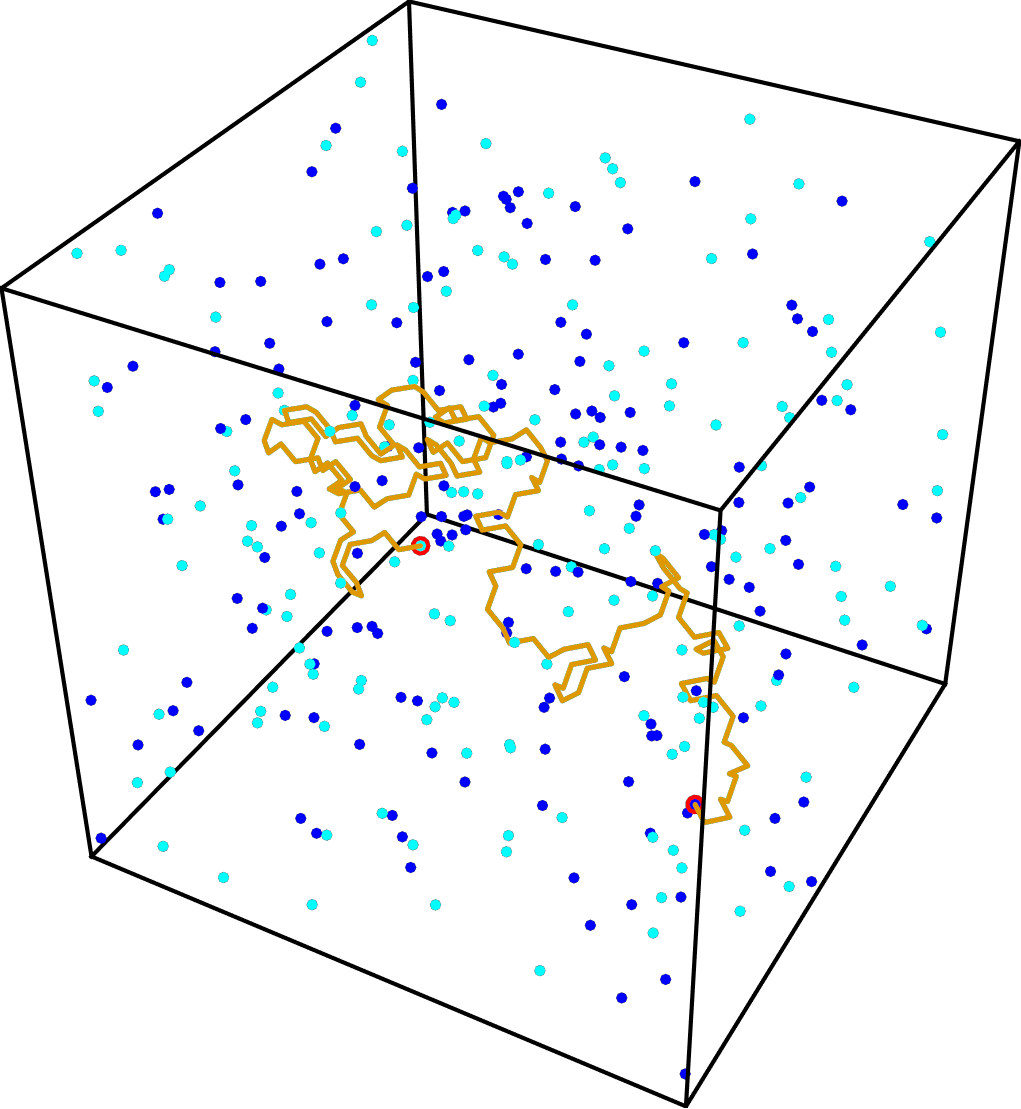}
    \caption{Example of a string segment in our numerical simulations of string percolation, connecting a monopole (blue points) and an antimonopole (cyan points), circled red. The initial monopole number density is taken to be $  n_M = 10^{-2} H_I^3$ and the random walk step size is $H_I^{-1}$.}
    \label{fig:random_walks}
\end{figure}

Our setup follows the algorithm of Refs.~\cite{Hindmarsh:1994ae,Strobl:1996yj}: the strings live on a tetrakaidecahedral lattice (a truncated octahedron with all edges equal), which is dual to a body-centred cubic (BCC) lattice broken up into tetrahedra. This lattice choice avoids the  ambiguities that arise on, e.g., cubic lattices when  joining string segments entering the same correlation volume, since  every tetrahedron only has one ingoing and one outgoing segment by construction. The lattice edge length corresponds to the correlation length $\xi$, which is $H_I^{-1}$ for inflationary production. We assign a random field value in the $\mathrm{U}(1)$ domain to points on the BCC lattice; this uniquely defines a self-interacting walk on the dual lattice, which is self-avoiding everywhere but which can close back on itself. Following \cite{Strobl:1996yj}, we do not perform the field value assignment over the whole grid at the start of the simulation, but only assign values to the vertices of tetrahedra being visited by the random walk. The grid gradually ``grows" with the string, which is both memory efficient and allows us to treat the lattice as being formally infinite. 

The simplest way to include monopoles is to set a fixed probability $p_M = \xi^3 n_M$ of the walk ending at each step. However, this ignores the fact that the distribution of monopole-antimonopole pairs is not quite uniform: pairs nucleated in the last few e-folds of inflation tend to be closer than average. To account for this, we first generate a uniform distribution of monopoles with density $n_M$ over the simulation volume, and then assign to each monopole the corresponding antimonopole based on the distribution in Eq.~\eqref{eq:Rdist}. When the walk encounters a monopole (antimonopole), it stops and continues in the opposite direction from its starting point until it encounters an antimonopole (monopole), and then terminates. The need to search the list of monopole-antimonopole positions at each step turns out to be the main limiting factor in the length of walks that we can easily simulate. Despite the grid being formally infinite, we only generate a realistic monopole-antimonopole distribution within a radius $D$ from the origin, which defines our simulation volume.

We perform $10^4$ walks for different monopole number densities scanning the range $p_M \in (10^{-6},10^{-1})$ as well as $p_M=0$, and extract the lengths of loops, finite segments and ``infinite" strings (those that leave the simulation volume). 
In the absence of monopoles, roughly one-third of walks close back on themselves and form string loops, with an average length of roughly $200 \xi$. As the monopole number density increases, both the fraction of loops and their average length decrease, because longer walks have a higher chance of encountering a monopole. 

\begin{figure}[t]
    \centering
    \begin{subfigure}[b]{0.49\textwidth}
         \centering
         \includegraphics[width=\textwidth]{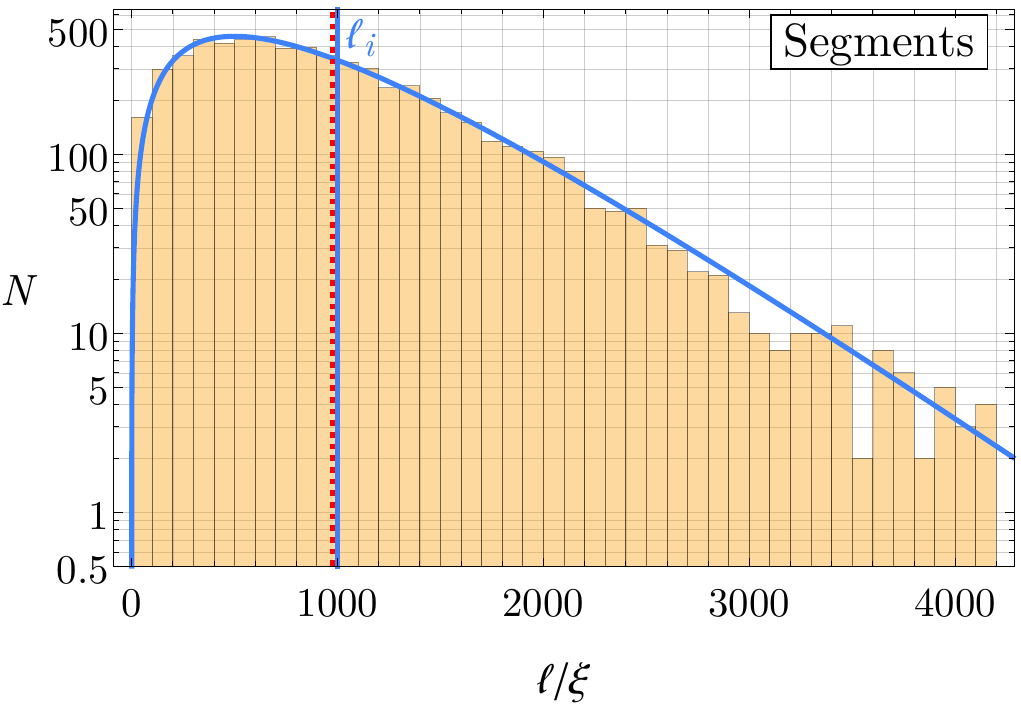}
    \end{subfigure}
    \hfill
    \begin{subfigure}[b]{0.49\textwidth}
          \centering
         \includegraphics[width=\textwidth]{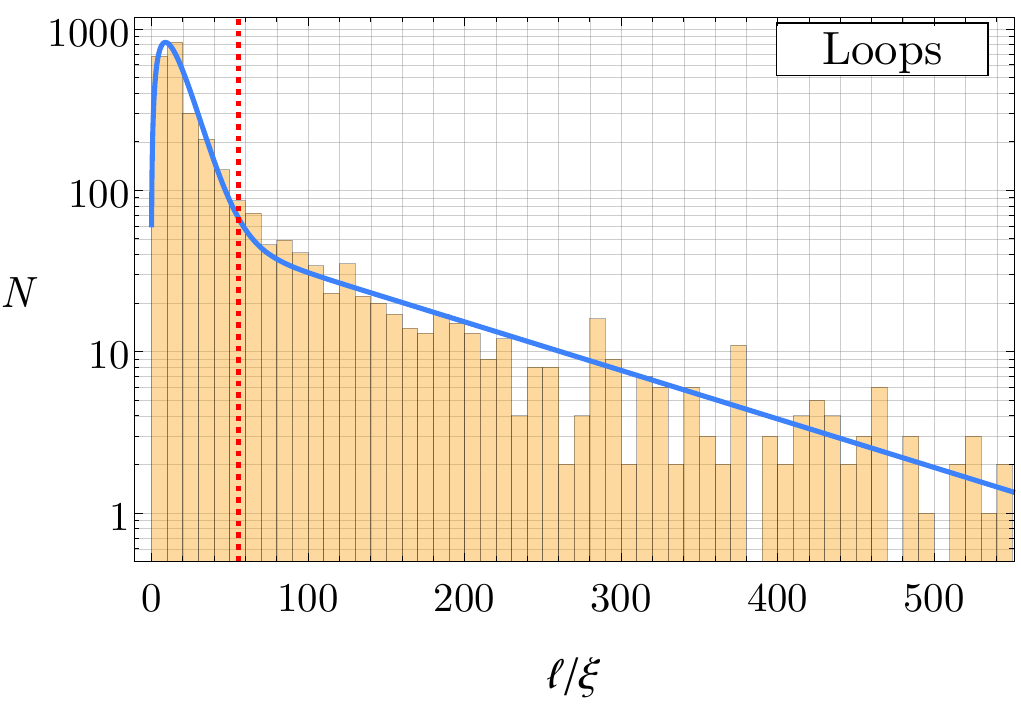}
    \end{subfigure}
    \hfill
    \caption{\textbf{Left:} Binned distribution of lengths of string segments for a monopole number density $n_M = 10^{-3} \xi^{-3}$ where $\xi$ is the correlation length, extracted from $10^4$ random walk simulations. The expected distribution, $P(\ell)= (\ell /\ell_i^2)\,e^{-\ell/\ell_i}$ with $\ell_i=\xi^{-2}/n_M$ (see Eq.~\eqref{eq:segmentdistribution}), is shown in blue; the expected average $\ell_i$ approximately coincides with the true average (red dashed line). \textbf{Right:} The same but for string loops. The blue line is now a best-fit distribution of the form $P(\ell)=b\ell \, e^{-a \ell/\xi} + d \,e^{-c\ell/\xi}$, with numerical values $a=0.112$, $b=8.25\cdot10^{-3} \xi^{-2}$, $c = 6.93\cdot10^{-3}$ and $d = 2.15\cdot10^{-3} \xi^{-1}$. At the value of the monopole density shown, around $28\%$ of strings form loops, which are typically much shorter than the segments.}
    \label{fig:ldist}
\end{figure}

Figure~\ref{fig:ldist} shows the typical length distribution for segments and loops. The distribution of segment lengths matches the expectation from Eq.~\eqref{eq:segmentdistribution} extremely well, suggesting that our realistic monopole distribution is well-approximated by a constant termination probability $p_M$, and nearby monopole-antimonopole pairs do not significantly bias this. Aside from the changing fraction of loops and long strings, simulations with other values of $n_M$ lead to the same qualitative conclusions.
For our purposes, the key conclusion of these simulations is that the probability of finding segments longer than the average is indeed exponentially suppressed in the presence of monopoles, even in the regime of very rare monopoles. This confirms and extends the behaviour found at high monopole densities in Ref.~\cite{COPELAND1988445}.

\section{Subsequent evolution} \label{sec:subsequent}

We now turn to the subsequent evolution of the finite-length segments and smaller loops, focusing on theories in which the typical segment length exceeds the Hubble distance at  formation. 

For some period after formation, while the Hubble distance remains sufficiently small, most Hubble patches contain no string endpoints. In such patches, the evolution resembles that of a standard string network: the system approaches a scaling regime in which the strings interact, intersect, and recombine as the Universe expands. Long strings continually lose length into loops, maintaining a configuration with roughly one Hubble length of string per Hubble volume. Meanwhile, in those patches that do contain string endpoints, the probability that the monopole and antimonopole endpoints of two segments meet is strictly less than, and not expected to be very close to, one. As a result, although some endpoints do recombine and annihilate, an order-one fraction will not. Those annihilations that do occur only affect the prefactors in our estimates such as Eq.~\eqref{eq:l0TM}, at the level of the other uncertainties.\footnote{This conclusion may change for global monopoles, which exert long-range forces and could recombine more efficiently.}

In Appendix~\ref{app:toymodel}, we use a simple analytic model of random recombination to argue that interactions among long but finite segments do not produce segments that are significantly longer than the mean and, in particular, do not regenerate infinite strings. 
String segments can also absorb loops, which can increase in their mean length. However, absorbing a loop generally leads to sub-horizon scale curvature that might be expected to be soon lost again in the scaling regime, and it does not affect the endpoint separation.

A potential complication arises from interactions between loops. Two loops can connect to form a single larger loop. In principle, sufficiently many repeated reconnections among many interlinked super-horizon loops could produce very large loops, which for practical purposes resemble infinite strings. Such loops would persist longer than the finite segments and would decay only through late-time quantum tunnelling.  
However, it would be surprising if arbitrarily large loops were produced, since a loop absorbed onto a finite string segment is effectively removed from the loop population (unless re-emitted). Long string segments dominate the length distribution at early times, and the probability for a sequence of percolating loops to avoid them is exponentially small. 

We also note that it appears more unlikely that large loops would be produced when the strings form by finite-temperature symmetry restoration, since finite-temperature breaking continues while the network approaches scaling. During this period the string density per Hubble patch decreases, reducing the number of loops available to interact and potentially percolate. For inflationary production the situation is more delicate, since breaking is only due to pre-existing monopoles, and one could imagine a complicated configuration of interlinked loops. Unfortunately, because the evolution depends sensitively on the detailed structure of the network, a statistical analysis analogous to that in Appendix~\ref{app:toymodel} does not appear informative. Numerical simulations of the network soon after formation, either field-theoretic or Nambu-Goto,  would be very useful, but we leave this for future work.

\subsection{Destruction of the network}

Assuming that loops do not percolate, and neglecting late-time vacuum-tunnelling breaking,  the dynamics described above continue until a typical long-string segment becomes sub-horizon. When this occurs depends on the endpoint separation $\ell_0$, rather than the initial contour length $\ell_i$ with the excess length shed as loops during the scaling regime. Once a segment is sub-horizon, it decays, with the monopole and antimonopole at its endpoints meeting and annihilating (we neglect the finite time that it takes for a sub-horizon segment to decay). Since even the longest initial loops are no larger than the initial segments, they also decay before this time.

During scaling, the endpoint separation grows with the scale factor $R(T)$ as $\ell(T)=\ell_0 (R(T)/R_i)$, where $R_i$ is the scale factor at the end of inflation. The evolution of the scale factor after inflation depends on the inflaton decay and the subsequent expansion history, introducing some model dependence into the analysis. We will typically assume that after the string network has formed, the Universe is in radiation domination (which corresponds to reheating just below $T=v$ in the inflationary production case), so the scale factor subsequently satisfies
\beq \label{eq:scalefactor}
\frac{R(T)}{R_{\rm RH}} \simeq \frac{T_{\rm RH}}{T} ~.
\eeq

We denote the temperature when the endpoints of a typical segment re-enter each other's horizons by $T_d$, and we restrict to theories in which this happens during radiation domination ($T_d \gtrsim {\rm eV}$). Using $H_d^{-1} = \ell_0 R(T_d)/R_i$, together with Eq.~\eqref{eq:scalefactor}, we obtain
\begin{equation}\label{eq:earlyTd}
    T_d = \frac{1}{\ell_0} \frac{M_{\rm Pl}}{T_{RH}} \, .
\end{equation}

As an illustrative example, consider thermal symmetry restoration with  $\sqrt{\mu}\simeq 10^{13}\,{\rm GeV}$ and let us require $T_d\sim {\rm MeV}$ to place the GW signals in the observable range.  Combining Eqs.~\eqref{eq:ell0} and \eqref{eq:earlyTd} then implies $m_M^2/\mu \gtrsim 400$. Inserting this value into Eq.~\eqref{eq:4} shows that quantum-tunnelling breaking is completely negligible at the time when the network is destroyed, so early-time breaking dominates.

\section{Breaking by vacuum tunnelling}\label{sec:late}

It is typically assumed that string breaking by quantum tunnelling at $T=0$ occurs at the rate given in Eq.~\eqref{eq:4}, with $\mu$ interpreted as the tension of a static straight string. 
In a realistic network, however, the energy density along a string is not uniform, but varies across the worldsheet. This variation can arise from regions of high curvature (such as kinks or cusps), from propagating modes, or from other dynamics. As a result, the local tension can exceed the analytical straight-string value, which we denote by $\mu_0$ in what follows. Since we consider theories with $m_M^2 \gg \mu$, even a modest enhancement of the local tension can dramatically increase the breaking rate, potentially making rare high-tension regions the dominant contributors.

We describe the distribution of tension over the network at time $t$ by a function $F(x,t)$, where $x = \mu/\mu_0$. This is normalised such that $\int_0^\infty dx\,F(x,t)=1$, so that $F(x,t) dx$ gives the fraction of the total string length with local tension in the interval $(x,x+dx)\mu_0$. 
Soon after formation the properties of the network are complicated and depend on the details of the initial conditions. If when first produced the strings are mostly straight, with curvature only on scales of order the correlation length $\xi$, $F(x,t)$ might be expected to initially be peaked near $x=1$, up to contributions from thermally excited Goldstone modes (see Section \ref{sec:reheat}). 
As the network evolves towards scaling, interactions and reconnections generate small-scale features, resulting in a distribution of tensions. Once scaling is reached, the tension distribution is expected to remain approximately stationary. We therefore drop the explicit time dependence when referring to the late-time distribution and simply write  $F(x)$. 

The distribution $F(x,t)$ depends on details of the network dynamics that are not analytically tractable, making its shape difficult to predict \textit{a priori}, even in the scaling regime. In the next subsection we extract a coarse-grained tension distribution from large field-theoretic simulations, which capture the evolution of the network from formation through to scaling. In general, we expect (and our simulations confirm) that $F(x,t)$ develops a tail extending to tensions above $\mu_0$. It is plausible that this tail extends to some maximum value $\mu_{\rm max}$ where the effective theory of the string breaks down and emission of e.g. radial modes becomes efficient, preventing larger tensions being reached.\footnote{The value of $\mu_{\rm max}$ could depend on the details of the theory. For strings arising from a two-stage symmetry breaking as discussed above, we can obtain a conservative upper bound by requiring that the local energy density does not exceed the monopole scale, implying $\mu_{\rm max} < \kappa \mu_0$.} The network is therefore destroyed when the total decay rate per Hubble time and Hubble length
\begin{equation}\label{eq:latetime}
    N_{H,\rm tot}(t) = H_d^{-2}(t) \int_0^\infty dx\, F(x,t) \Gamma(x \mu_0) \approx 1 \, .
\end{equation}
reaches order-one, where the decay rate $\Gamma(\mu)$ is treated as a function of the local tension $\mu$.  

The local tension on the string worldsheet is generally non-uniform and time dependent, with large gradients potentially developing near sharp features such as cusps. Such inhomogeneities modify the standard bounce-action calculation for the decay rate. In principle, the action depends on the detailed dynamics of the worldsheet degrees of freedom responsible for the increased energy density, and different sources of tension enhancement need not affect the decay rate in identical ways. Consequently, one cannot in general obtain the decay rate by simply substituting the local tension into Eq.~\eqref{eq:4}. Nevertheless, it is reasonable to expect that the local tension provides a useful proxy for the leading effect on string breaking. Since a fully consistent calculation of the bounce action including worldsheet dynamics is beyond the scope of this work, we employ this simplified tension-based prescription as a practical estimate of the decay rate.

Within this approximation, it is useful to consider a long string with uniform tension $\mu_0$ everywhere except near some feature, where it is enhanced over a length scale $\lambda$. As discussed in Section \ref{sec:reheat}, the instanton associated with the baseline tension has radius  $r_M = m_M/\mu_0 = \kappa/m_M$. In the situations of interest the tension enhancement is at most a factor of a few, so even in a high-tension region the size of the instanton remains comparable to $r_M$. The  bounce action then depends on the relative sizes of $\lambda$ and $r_M$. 
If $\lambda \gg r_M$, the instanton probes a nearly uniform region, and replacing $\mu_0$ by an effective local tension provides a reasonable estimate of the decay rate. If $\lambda \sim r_M$, the computation becomes difficult; the decay rate is likely increased, but Eq.~\eqref{eq:4} should not be expected to be quantitatively reliable. If $\lambda \ll r_M$, the instanton effectively averages over the feature, and its impact on the decay rate is parametrically suppressed. 

Motivated by this picture, we estimate the breaking rate using Eq.~\eqref{eq:4} with $\mu$ averaged over regions of size $l\simeq r_M$, which captures both the $\lambda \gg r_M$ and $\lambda \lesssim  r_M$ regimes. To the level of accuracy that we can realistically achieve, this averaging procedure provides a reasonable estimate of the impact of regions of locally enhanced tension.

\subsection{Tension distribution from numerical simulations}

To obtain the local tension distribution and assess its impact on the network's lifetime, we use numerical simulations of critically coupled Abelian-Higgs strings. In these simulations, we evolve the gauge field and a complex scalar with a symmetry-breaking potential on a discrete spatial lattice using a standard leapfrog algorithm in a radiation-dominated background (technical details are given in Appendix~\ref{app:furthersim}).

Starting from suitable random initial conditions, a string network forms and evolves towards the scaling solution. However, such simulations cannot access the physical hierarchy between the Hubble distance $H^{-1}$ and the string thickness, of order $m_r^{-1} \sim v^{-1}$, because computational constraints restrict us to grids containing roughly $3000^3$ lattice points. In particular, the lattice spacing $\Delta$ must satisfy $\Delta m_r\lesssim 1/2$ to evolve the string cores as in the continuum limit, while the grid must contain at least a few Hubble patches to capture the overall structure of the network. 
As a result, we are limited to  $\log(m_r/H)\lesssim 7$, far below the values $\log(m_r/H)\gtrsim 60$ relevant at the time of GW emission.\footnote{The results shown are from ``fat string'' simulations, in which action is modified so that the comoving core size is time-independent; this allows longer evolution in cosmological time and a closer approach to the attractor. We have verified that simulations with the physical action show the same key features, including the same exponential fall-off in the frequency of large tensions, with fluctuations depending on the initial conditions.}

Nevertheless, we expect that properties of the network, in particular the distribution of string tensions, extracted at small scale separations still capture the broad features in the physical regime. A detailed analysis of the Abelian-Higgs network can be found in \cite{usfuture}; related simulations of gauge strings include \cite{Bevis:2006mj,Hindmarsh:2008dw,Hindmarsh:2017qff}, and similar analyses for global strings can be found in Refs.~\cite{Gorghetto:2018myk,Gorghetto:2020qws}.  

\begin{figure}[t]
    \centering
    \begin{subfigure}[t]{0.49\textwidth}
         \centering
         \includegraphics[width=\textwidth]{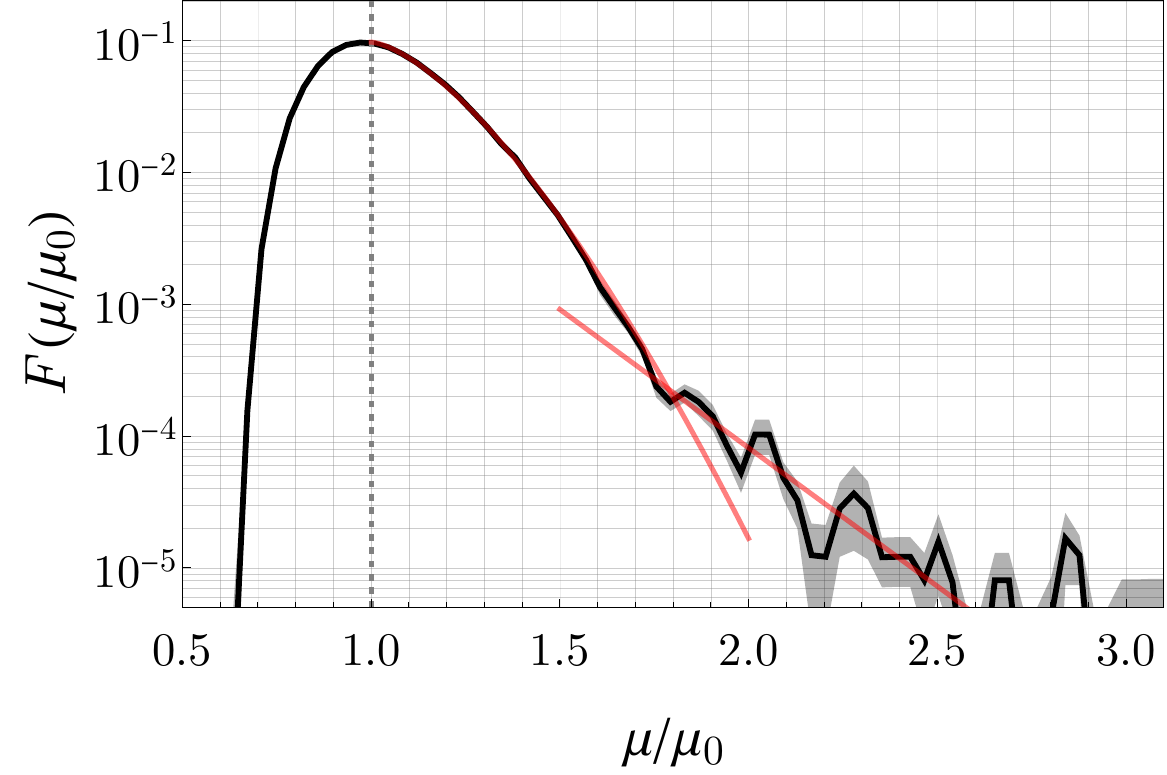}
    \end{subfigure}
    \hfill
    \begin{subfigure}[t]{0.49\textwidth}
          \centering
         \includegraphics[width=\textwidth]{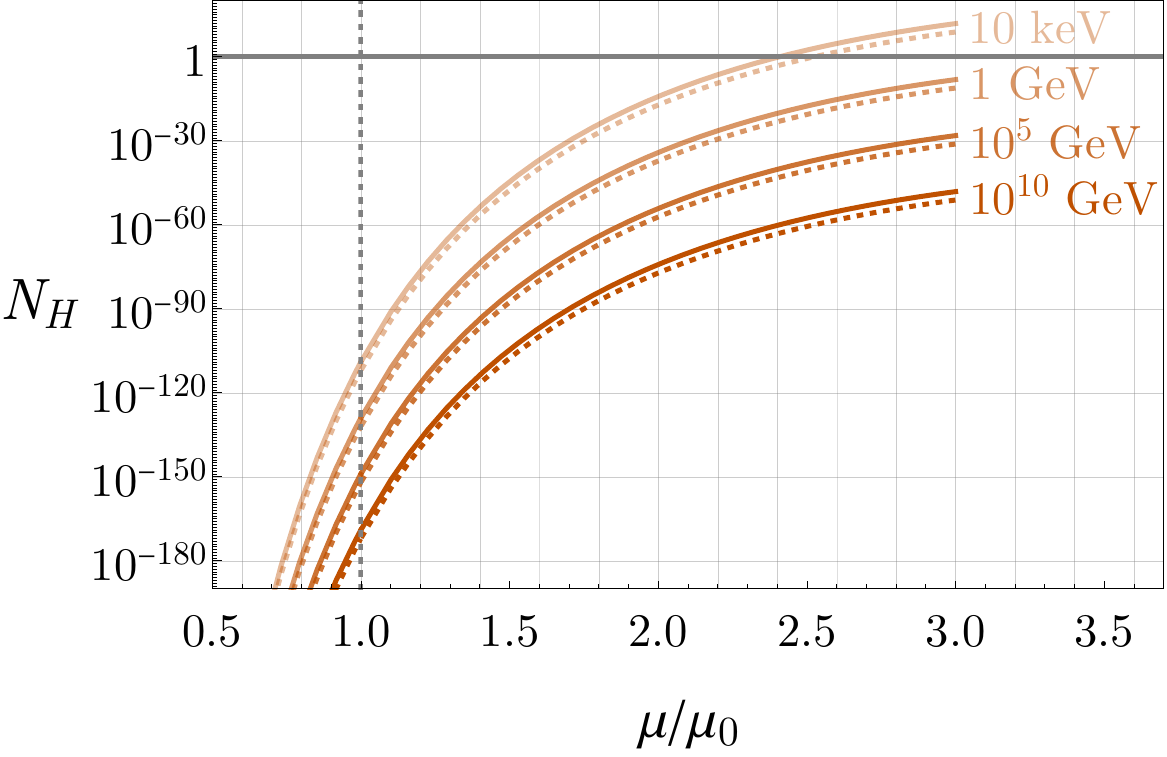}
    \end{subfigure}
    \hfill
    \caption{\textbf{Left:} Distribution of tensions for an Abelian-Higgs critical string network (with straight string tension $\mu_0 = 2\pi v^2$) extracted from numerical simulations, with statistical uncertainties (shaded). The red lines show the best-fit behaviour in the regions $\mu < 1.7\mu_0$ and $\mu > 1.7\mu_0$ respectively. \textbf{Right:} The number of breaking events per Hubble time and Hubble length due to string regions with tension between $\mu$ and $\mu+d\mu$ at different cosmological times in radiation domination, based on the fit of Eq. \eqref{eq:latetimeFit} extrapolated to $\mu/\mu_0 \simeq 3$. Breaking dominantly happens at rare high-tension regions, since the enhancement of the breaking from the increased tension outweighs the rarity of such regions. We fix $\kappa = 144$, and plot the results at different values of Hubble (labelled by cosmic temperature) and for two values of tension, $G_{\rm N}\mu_0 = 10^{-7}$ (thick lines) and $G_{\rm N}\mu_0 = 10^{-10}$ (dashed lines).}
    \label{fig:mudist}
\end{figure}

To extract the local string tension, we first identify the string cores on the lattice. We then compute the total field energy within a cube of side $l$, using a screening criterion to subtract the contribution from free gauge-boson and radial mode waves. We also determine the length $\ell_k$ of the string segment inside the cube and its average Lorentz factor $\gamma_k$. Then, the (rest frame) tension of the string segment is schematically given by
\begin{equation}\label{eq:tensionmethod}
    \mu_k = \frac{l^3}{c_{\rm sc} \gamma_k \ell_k} \sum_{i} \rho_i \, ,
\end{equation}
where $k$ labels all possible cubes, and the sum in $i$  runs over all string core points within a cube. The coefficient $c_{\rm sc}$ is an $\mathcal{O}(1)$ number correcting for the string energy located outside the very core of the string. Based on the radial profile of a straight-string, we fix $c_{\rm sc} = 0.85$. Repeating this procedure across the whole network at different times yields the distribution $F(\mu/\mu_0,t)$, coarse-grained over the length scale $l$.

In addition to the need to extrapolate from the simulated to the physical scale separation, another major source of uncertainty is the choice of averaging scale $l$. As discussed above, $l$ should be comparable to $r_M$, but the combination $r_M m_r$ depends on $\kappa$. In principle, one should extract results for different values of $l$ to cover the relevant range of $r_M$. In practice, $l$ must also be large enough to encompass several lattice sites for a reliable estimate of $\mu$.  

Given these considerations, we extract $F$ using $l=2r_M$ for the representative value $\kappa = 64$, which corresponds to $l=\sqrt{2\kappa/\pi}\,m_r^{-1} \approx 6.4 \, m_r^{-1}$. We expect this choice to provide a reasonable estimate of the tension distribution relevant for  $\kappa \lesssim 250$. For larger $\kappa$ the averaging region should be correspondingly larger, so the distributions we obtain are less reliable, although we expect the qualitative behaviour to persist.

Another potential source of uncertainty is gravitational backreaction, which is not included in our simulations. It has been argued, mostly based on Nambu-Goto simulations (see e.g. \cite{Blanco-Pillado:2018ael,Wachter:2024aos}), that gravitational backreaction tends to smooth out small-scale features on strings, thereby modifying the tension distribution. In particular, this smoothing would  reduce the fraction of regions with very high tension, causing $F(\mu)$ to fall off more rapidly at large $\mu$. It is difficult to estimate how backreaction would affect our results, and we leave a detailed study to future work.

The tension distribution at the end of the simulations (at $\log(m_r/H) = 6.5$) is shown in the left panel of Figure~\ref{fig:mudist}. As expected, the distribution is peaked around $x = 1$ and falls off exponentially for $x > 1$. In the range $1 < x \lesssim 1.7$, it is well fit by
\begin{equation}\label{eq:tailmu}
    F(x) \simeq A e^{-B(x-1)^{3/2}}\, , \quad A = 0.097\, , \quad B= 8.67\, .
\end{equation}
For $x \gtrsim 1.7$, the distribution decreases more slowly, approximately scaling as $\sim e^{-x}$. We believe that this slower fall-off is likely an artifact of simulations and would not persist in the physically relevant regime. 
The initial conditions we use produce a relatively dense string network at early times, before it relaxes to scaling; regions of high tension may therefore be leftover remnants of this early era when there are frequent string intercommutations and tend to diminish as the network evolves. Moreover, the small values of $\log(m_r/H)$ accessible in simulations mean that, at any moment, the fraction of string length involved in intersections (of order $\left(H/m_r\right)^2$, since each Hubble-length of string intersects another once per Hubble time for a typical duration $m_r^{-1}$ over a distance $m_r^{-1}$) is much larger than in the physical regime where $\log(m_r/H)\gtrsim 70$. 
In Appendix \ref{app:furthersim}, where $F(x,t)$ is plotted at earlier simulation times, we indeed observe a more prominent high-tension tail that gradually evolves towards the behaviour described in Eq. \eqref{eq:tailmu}.

It is therefore reasonable to take the distribution in Eq. \eqref{eq:tailmu} as an estimate of the scaling distribution for $x > 1$. As mentioned above, on physical grounds we also expect that there might be a sharp cut-off in the tension distribution at some maximum value $\mu_{\rm max}$, above which $F(x)$ should be strongly suppressed. This feature is not seen in simulations, perhaps because of the transient tail discussed above and because the statistics of the sampling at large $x$ is limited (making our extraction of the distribution unreliable for $x \gtrsim 3$). In what follows, we impose such a cut-off by hand. 

Plugging the resulting distribution into Eq.~\eqref{eq:latetime}, we can compute the total number of decays per Hubble patch per Hubble length, $N_{H,\rm tot}$:
\beq\label{eq:latetimeFit}
N_{H,\rm tot} = \int_1^{x_{\rm max}} dx\, N_H (x) = H_d^{-2}(t) \frac{A \mu_0}{2\pi} \int_1^{x_{\rm max}} dx\, x \exp{\left[-B(x-1)^{3/2} - \frac{\pi \kappa}{x} \right]}  \, ,
\eeq
where the anticipated cut-off appears as the upper limit of the integral $x_{\rm max}=\mu_{\rm max}/\mu_0$. The integrand $N_H(x)$ represents the expected number of breaking events contributed by string regions with tension in the interval $(x, x+dx)$. It increases monotonically up to a peak at some value $x_{\rm s}$, beyond which the rarity of high-tension regions dominates over the enhancement in the decay rate. Given the form of $N_H(x)$, the integral in Eq.~\eqref{eq:latetimeFit} is, to the accuracy that we require, well approximated by
\begin{equation}\label{eq:latetimeNH}
   N_{H, \rm tot} \simeq 
   \begin{cases} 
   N_H(x_{\rm s})  \qquad  & {\rm if }~x_{\rm s}<x_{\rm max}\\
   N_H(x_{\rm max}) & {\rm otherwise} ~.
   \end{cases}
\end{equation}

We plot $N_H(x)$ in the right panel of Figure~\ref{fig:mudist} for several values of the Hubble parameter (i.e., at different times in the cosmological history) with fixed $\kappa=144$. In doing so, we extrapolate the fit of Eq.~\eqref{eq:tailmu} out to $\mu/\mu_0 =3$, beyond the values where it matches the simulation data. Crucially, over this range the integrand in Eq.~\eqref{eq:latetimeFit} continues to increase, implying that breaking is dominated by rare high-tension regions. In other words, when $N_H(x_s) \sim 1$, we have $N_H(1) \ll 1$; 
a constant-tension estimate with $\mu=\mu_0$  would therefore incorrectly suggest that the network behaves like a stable network at that time, whereas it is actually already being destroyed by its high-tension regions.

It is straightforward to determine the peak of $N_H (x)$  based on Eq. \eqref{eq:tailmu}. Numerically, we find that $x_{\rm s}$ grows with $\kappa$: for example, $x_{\rm s} \simeq 3.24$ when $\kappa = 64$ and $x_{\rm s} \simeq 5.39$ when $\kappa = 250$. These values lie well beyond the range of $x$ for which the fit in Eq. \eqref{eq:tailmu} agrees with the simulation data,  so we cannot be confident that such an extrapolation is reliable, and it might be that the physical cut-off $x_{\rm max}$ appears earlier. 
For this reason, in the phenomenological analysis that follows we take $x_{\rm max} \approx 1.75< x_{\rm s}$, which is the largest value for which the simulation data matches the fitted distribution. Our results should therefore be interpreted as a conservative lower bound on the breaking rate, based on the range of tensions directly probed in simulations. 

With the assumptions above, and assuming radiation domination, breaking by vacuum tunnelling would lead to the network being destroyed at a temperature~\footnote{As discussed above, in Eq.~\eqref{eq:lateTd} we assume $F$ is evaluated after averaging over distances of order $r_M$, such that the decay rate can be approximated by $\Gamma(\mu)$ from Eq.~\eqref{eq:4} with $\mu$ the local tension.}
\begin{equation}\label{eq:lateTd}
    T_d = \sqrt{M_{\rm Pl}} \left( \frac{ x_{\rm max}\mu_{0} \, F(x_{\rm max})}{2\pi}\right)^{1/4} \exp{\left(-\frac{\pi \kappa}{4  x_{\rm max}} \right)} \, .
\end{equation}
Even as a lower bound, the temperature obtained from Eq. \eqref{eq:lateTd} is exponentially larger than the value predicted by the constant-tension estimate of Eq. \eqref{eq:TdestQ}, although for $\kappa \gg 1$ it remains smaller than the destruction temperature resulting from early-time breaking, which scales as $\sim e^{\sqrt{\kappa}}$.

\section{The parameter space of metastable strings} \label{sec:signals}

The processes described in the previous sections destroy a metastable string network much earlier than previous estimates based on uniform-tension quantum tunnelling predict, with important phenomenological consequences (we re-emphasise the caveat that we assume loops do not re-percolate, although this does not affect the enhancement of the vacuum-breaking rate from high tension regions).  
The potential observational signals from metastable strings depend sensitively on the time at which the network decays, characterised by the cosmic temperature at this point $T_d$ given in Eqs.~\eqref{eq:earlyTd} and \eqref{eq:lateTd}. The primary observational signature we consider, and the one that has received the most attention in recent years, is the stochastic gravitational wave background (SGWB) produced by a metastable string network. In this case, earlier network destruction alters the SGWB's dependence on the hierarchy between the monopole mass  and the string tension $\kappa$. As we will show, the main consequence is a shift of the low-frequency cutoff of the signal to higher frequencies for a given $\kappa$, such that a much larger $\kappa$ is required to account for the NANOGrav signal.

For a network produced by inflationary fluctuations, the only relevant early-time breaking process is the pre-existing population of monopoles. For a network produced by thermal restoration, the dominant early-time process is also the pre-existing monopoles; finite-temperature enhanced decays are less important across the relevant parameter space, although (as discussed in Section~\ref{sec:reheat}) they are more model-independent and rely on fewer assumptions. 
\begin{figure}
    \centering
    \includegraphics[width=0.7\linewidth]{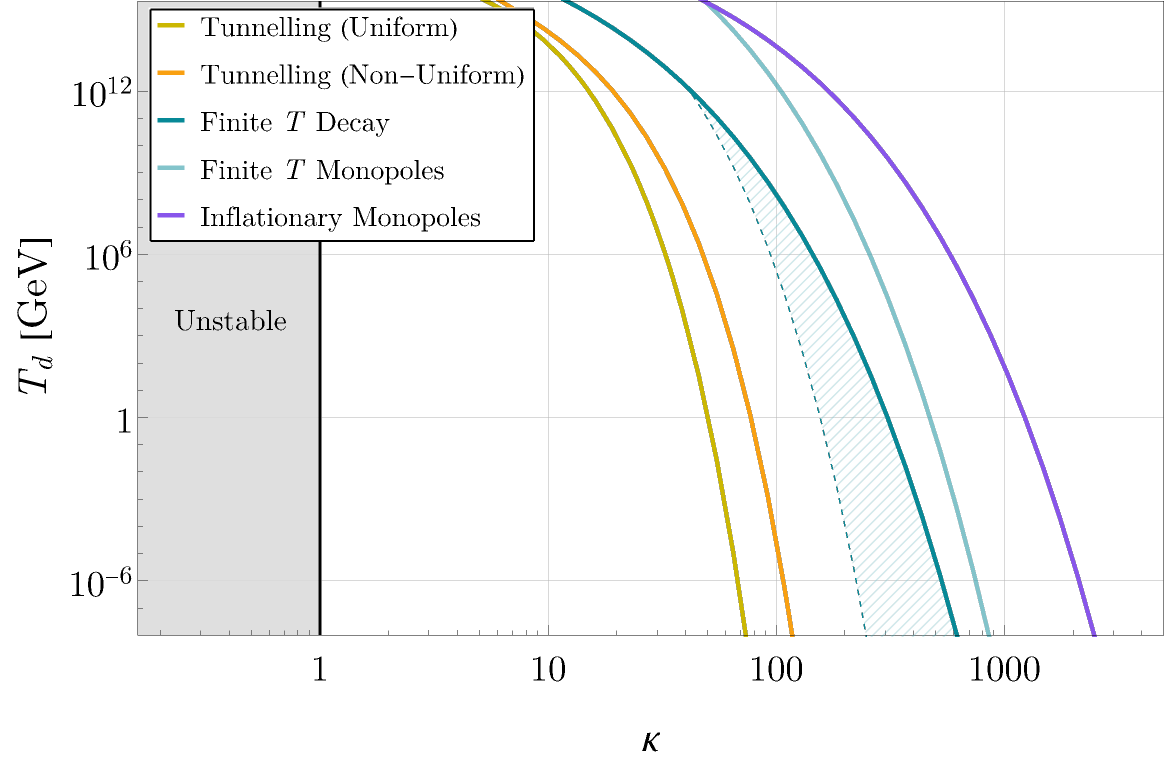}
    \caption{Dependence of the temperature $T_d$ on the ratio of scales $\kappa$, for the various early and late-time destruction mechanisms discussed in this work. We fix $G_{\rm N}\mu = 10^{-7}$, $\Gamma_g = 50$ and (where relevant) $T_{RH} \simeq v$, $H_I/(2\pi) \simeq v$. The blue shaded region shows how our prediction for $T_d$ is affected by taking a more conservative estimate for the finite-temperature decay rate, as explained at the end of Section \ref{subsec:finiteT}.}
    \label{fig:Tdall}
\end{figure}
To compare the relevance of early- and late-time breaking, we must account for the fact that the Hubble time is much larger at later epochs. To make this comparison, in Figure \ref{fig:Tdall} we plot the cosmic temperature at which the network decays, $T_d$, as a function of $\kappa$, using Eqs.~\eqref{eq:earlyTd} and \eqref{eq:lateTd} for early and late-time breaking respectively. We also show the result obtained assuming late-time breaking with a uniform tension. In this, for thermal production we take the minimum allowed values of the reheating temperature $T_{\rm RH}=v$, and for inflationary production we fix $H_I/(2\pi)=v$. This leads to lower bounds on the required values of $\kappa$ for a given $T_d$ for breaking by pre-existing monopoles in the two cases.

We see that late-time breaking is subdominant over the relevant parameter space, and carries the largest uncertainties due to the potential effects of gravitational backreaction and limitations of the simulations. As expected, finite-temperature decay is likely somewhat less important than pre-existing monopoles, and for inflationary production even larger $\kappa$ are required for a given $T_d$.

For simplicity, in the following we focus on networks that decay before matter-radiation equality ($T_d \gtrsim {\rm eV}$), which approximately corresponds to $\kappa \lesssim 3000$ for $H_I/(2\pi)\simeq v$ or $T_{\rm RH}\simeq v$. 

\subsection{Gravitational Wave Spectrum}

The stochastic gravitational wave background produced by a network of cosmic strings has been extensively studied for many years \cite{Vachaspati:1984gt,Damour:2000wa,Damour:2004kw,Auclair:2019wcv}. The continuous emission of GWs from loops during radiation domination produces a characteristic spectrum that is predicted to be approximately flat over frequencies spanning several orders of magnitude \cite{Vilenkin:2000jqa}. For stable strings, the primary physical parameter controlling the spectrum is the string tension $\mu$, conventionally quoted as the dimensionless quantity $G_{\rm N}\mu$. However, the detailed form of the spectrum also depends on the choice of the specific model used to describe the network, and on several $\mathcal{O}(1)$ parameters calibrated from simulations (see \cite{Auclair:2019wcv} for a comprehensive review). 

The intensity of a GW signal is typically expressed as the energy density in gravitational waves relative to the critical density, denoted $\Omega_{\rm gw}$. Following \cite{Auclair:2019wcv}, the height of the spectral plateau is given by
\begin{equation}\label{eq:omegagw}
    \Omega_{\rm gw}^{\rm plateau} \simeq \frac{128\pi}{9} B \Omega_r \left( \frac{G_{\rm N}\mu}{\Gamma_g}\right)^{1/2} \, ,
\end{equation}
where $\Gamma_g$ is the radiation efficiency into GWs, typically taken to be $\sim 50$ for loops, $\Omega_r \simeq 8.97 \, \times \, 10^{-5}$ is the radiation density \cite{Planck:2018vyg} and $B \simeq 0.18$ is a numerical parameter of the BOS model \cite{Blanco-Pillado:2013qja} extracted from Nambu-Goto simulations.

The SGWB produced by metastable strings closely resembles that from stable strings, with two crucial differences. 
First, the network decays away at some point and stops emitting gravitational waves, so the spectral plateau has a low-frequency cutoff around a characteristic frequency $f_{\rm low}$. Second, as the network is destroyed the SGWB spectrum receives contributions from oscillating sub-horizon segments in addition to loops. The resulting features have been studied in detail in Ref.~\cite{Buchmuller:2021mbb}. Below we simply summarise a few key results and adapt them to our purposes.

The spectrum is now determined by two physical parameters: the tension $\mu$ and the separation of scales $\kappa$. The network begins to decay at cosmic time $t_d$, corresponding to temperature $T_d$. Afterwards, sub-horizon segments continue to oscillate and lose length to loops, meaning that the emission of GWs will continue until some later time $t_e$, when the strings have fully disappeared. 

The time at which the emission of gravitational waves ends is defined as
\begin{equation}
    t_e = \left(\frac{2}{\tilde{\Gamma}_g \, G_{\rm N}\mu} \right)^{1/2} t_d \, ,
\end{equation}
where $\tilde{\Gamma}_g$ is the radiation efficiency of oscillating segments. In general, $\tilde{\Gamma}_g \sim 4 \ln{\gamma_0^2}$, where $\gamma_0$ is the largest Lorentz factor achieved by the string endpoints (monopoles) as they oscillate, which is model dependent. If the monopoles experience significant friction, they might not oscillate at all and simply annihilate, such that $t_e \sim t_d$. Here we follow \cite{Buchmuller:2021mbb} and take $\tilde{\Gamma}_g \simeq \Gamma_g \simeq 50$. 

The present-day GW density produced by metastable string loops per logarithmic frequency interval $f > f_{\rm low}$ is given by
\begin{equation}\label{eq:fullspectrum}
    \Omega_{\rm gw} (f) \simeq \frac{128\pi}{9} B \Omega_r \left( G_{\rm N}\mu\right)^2 \frac{\Gamma_g}{\zeta(4/3)} \sum_{k =1}^{k_{\rm max}} {k^{-4/3} \left[\frac{2k}{f}\sqrt{\frac{2 H_r}{ t_e}} + \Gamma_g G_{\rm N}\mu \right]^{-3/2}}
\end{equation}
where $k_{\rm max}$ is the maximum harmonic excited as the strings oscillate, and we focus on the emission from string cusps (hence the $\sim k^{-4/3}$ scaling). The low-frequency cutoff in the GW plateau is
\begin{equation} \label{eq:flowdef}
    f_{\rm low} = \frac{4}{\Gamma_g \, G_{\rm N}\mu} \sqrt{\frac{H_r}{2 t_e}} = T_d \left( \frac{2}{\Gamma_g \, G_{\rm N}\mu} \right)^{3/4} \sqrt{\frac{H_r}{M_{\rm Pl}}} \, ,
\end{equation}
with $H_r \equiv H_0 \sqrt{\Omega_r}$. Substituting $\Omega_r = 8.97 \, \times \, 10^{-5}$ and $H_0 = 10^{-33} \, {\rm eV}$ yields the convenient expression
\begin{equation}
    f_{\rm low} \simeq 9.7 \times 10^{-4} \left( \frac{\rm Hz}{\rm GeV}\right) \left( \frac{50}{\Gamma_g}\right)^{3/4} \left( \frac{10^{-7}}{G_{\rm N}\mu}\right)^{3/4} \, T_d \, .
\end{equation}
Considering only quantum tunnelling (Eq.~\eqref{eq:TdestQ}), for the benchmark values $G_{\rm N}\mu = 10^{-7}$, $\Gamma_g = 50$ and $\kappa = 64$ we obtain $f_{\rm low} \sim 10^{-8} \, {\rm Hz}$, in agreement with the results of Ref.~\cite{Buchmuller:2021mbb}.

The spectrum also has a high-frequency cutoff $f_{\rm high}$, corresponding roughly to the time when the network first begins to emit GWs. Expressing this cutoff in terms of the cosmic temperature at this time, $T_i$, and assuming radiation domination throughout,
\begin{equation}
    f_{\rm high} \simeq 3.3 \times 10^{-2} \left( \frac{\rm Hz}{\rm GeV}\right) \left( \frac{50}{\Gamma_g}\right) \left( \frac{10^{-7}}{G_{\rm N}\mu}\right) \, T_i \, .
\end{equation}
For strings formed by finite-temperature symmetry breaking, $T_i = T_\star \simeq v$, which leads to $f_{\rm high} \sim 10^{13} \, {\rm Hz}$ for the usual benchmark $G_{\rm N}\mu \sim 10^{-7}$. For inflationary fluctuations, the form of the spectrum at frequencies corresponding to very early times depends on the details of reheating. In the following, we will take $T_i = T_{RH} \simeq v$, consistent with our earlier assumptions. In practice, this typically does not affect the SGWB signal in frequencies probed by near-future observations unless reheating is very slow.

The additional contribution to the GW spectrum from short segments was also studied in Ref.~\cite{Buchmuller:2021mbb}. Provided $\tilde{\Gamma}_{g} \gtrsim \Gamma_g$, this is found to be subdominant compared to the emission from loops, and the intensity of the signal $\Omega_{\rm gw}^{\rm plateau}$ remains well approximated by Eq.~\eqref{eq:omegagw}.
\begin{figure}[h!]
    \centering
    \includegraphics[width=0.8\linewidth]{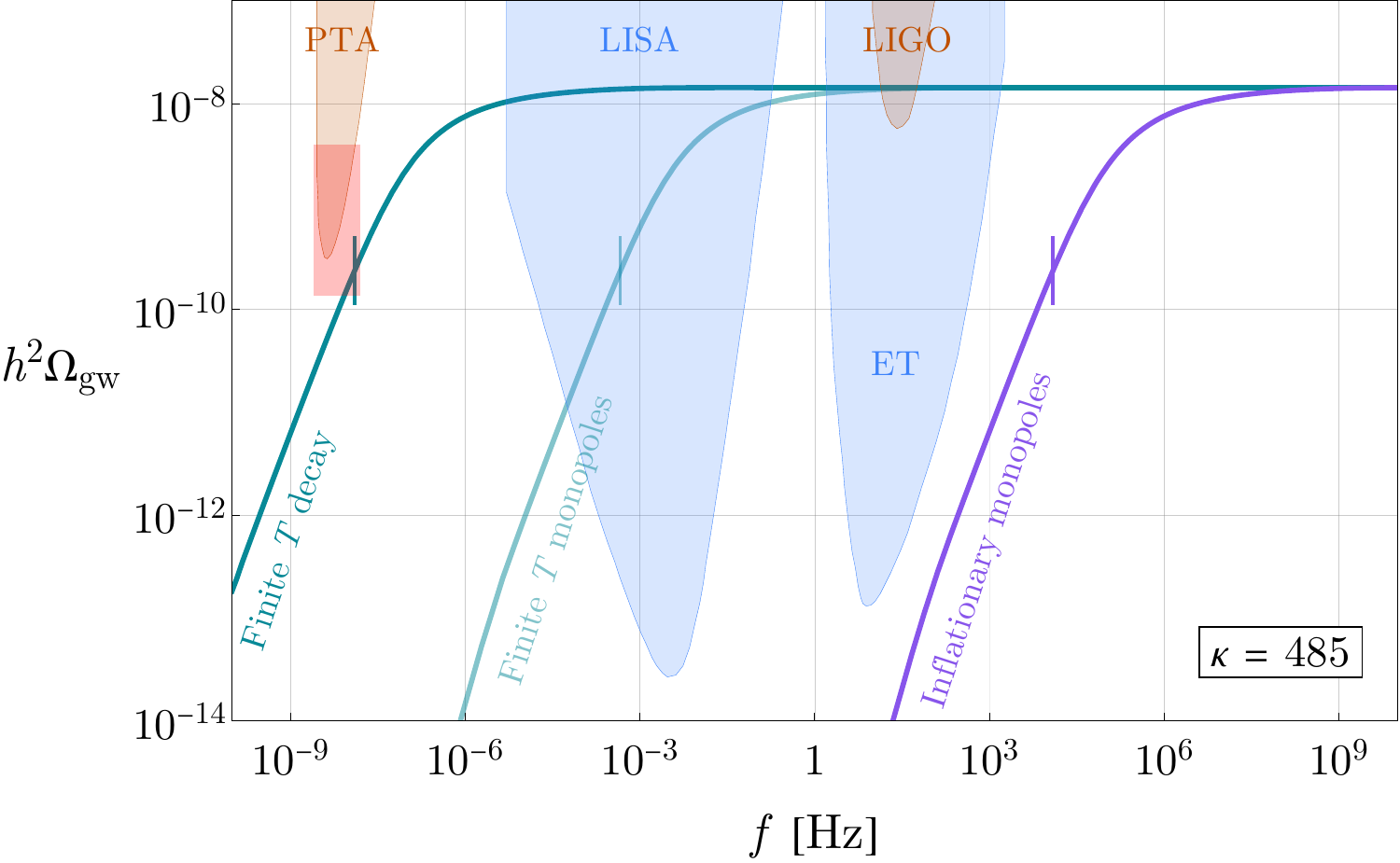}
    \caption{The GW spectrum from a metastable cosmic string network, rescaled by $h^2 = (0.68)^2$, compared to the design sensitivities of current (shaded orange) and future (shaded blue) GW detectors. The red rectangle indicates the approximate location of the stochastic GW signal observed by PTA. We fix $\kappa = 485$, $G_{\rm N}\mu = 10^{-7}$, $\Gamma_g = 50$, and (where relevant) $T_{RH} \simeq v$, $H_I/(2\pi) \simeq v$. The spectra shown correspond to the different network formation and early-time breaking mechanisms considered in this work. For networks formed by finite-temperature restoration, enhanced breaking due to finite-temperature effects leads to the spectrum labelled ``Finite T decay'' (dark blue-green line), while including the additional effect of thermally produced monopoles leads to earlier network destruction (``Finite T monopoles'', light blue line). Taking a more conservative estimate of the decay rate at finite T, the network decays at much later times, and we need a smaller value of $\kappa \simeq 200$ to match the PTA signal (see Figure~\ref{fig:frequencies}). For a network formed by inflationary fluctuations, monopoles produced during inflation result in the spectrum labelled ``Inflationary monopoles'' (purple line); in this case much larger $\kappa\simeq 2000$ is required to match the PTA signal. The ticks along the curves correspond to the location of $f_{\rm low}$, defined in Eq.~\eqref{eq:flowdef}.}
    \label{fig:GWspectrum}
\end{figure}
\begin{figure}[t]
    \centering
    \includegraphics[width=0.8\linewidth]{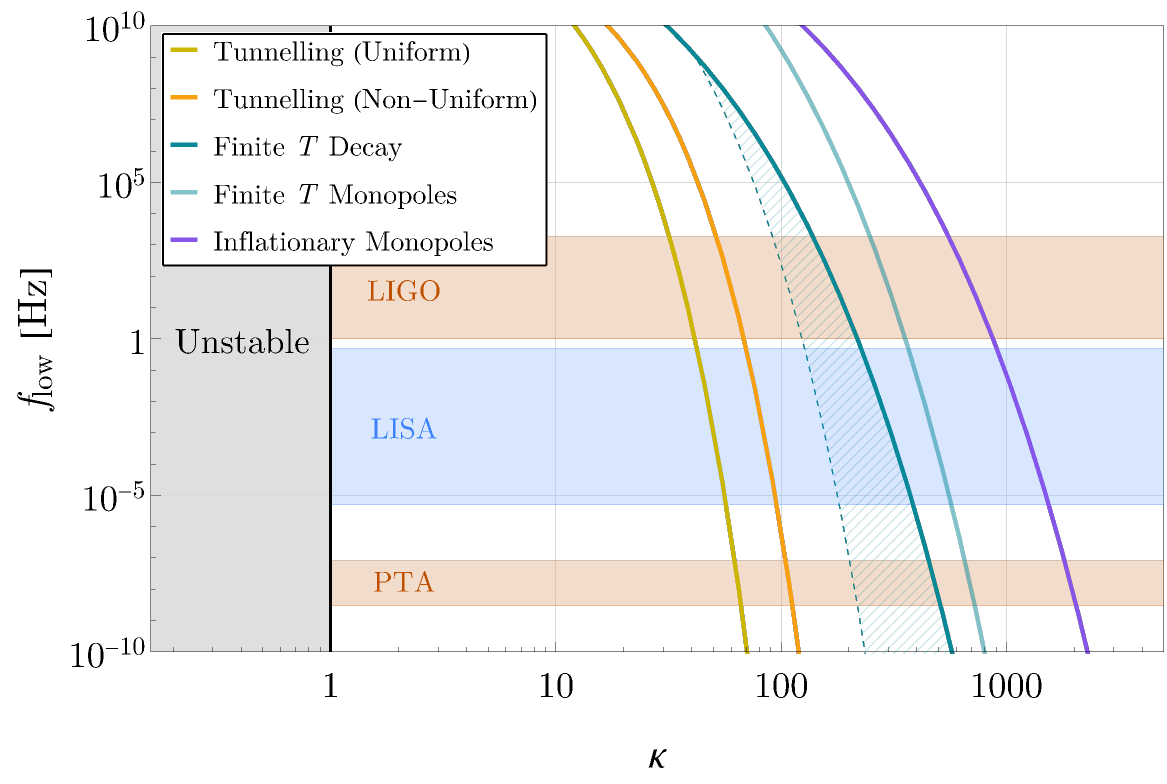}
    \caption{Dependence of the cutoff frequency $f_{\rm low}$ on the ratio of scales $\kappa$, compared with the approximate sensitivity bands of current (shaded orange) and future (shaded blue) GW detectors, for the various destruction mechanisms discussed in this work. We fix $G_{\rm N}\mu = 10^{-7}$, $\Gamma_g = 50$ and (where relevant) $T_{RH} \simeq v$, $H_I/(2\pi) \simeq v$. The shaded region between the thick and dashed blue-green lines shows how our prediction for $f_{\rm low}$ is affected by taking a more conservative estimate for the finite-temperature decay rate, as explained at the end of Section \ref{subsec:finiteT}. }
    \label{fig:frequencies} 
\end{figure}

Recently, a GW signal in the ${\rm nHz}$ band detected by pulsar timing arrays (PTA) has been used to place stringent constraints on the plateau, $\Omega_{\rm gw}^{\rm plateau}$, produced by a network of cosmic strings,\footnote{Note that these constraints rely on Nambu-Goto modelling of GW emission from local strings, which is not supported by field-theoretic simulations. Using field theory results weakens the bound on $G_{\rm N}\mu$ by orders of magnitude, see e.g. \cite{Baeza-Ballesteros:2024sny,Kume_2024}.} implying $G_{\rm N}\mu \lesssim 10^{-10}$ \cite{NANOGrav:2023newphysics}. 
The PTA signal has sparked significant interest in metastable string networks, see e.g. \cite{Buchmuller:2020lbh,Masoud:2021prr,Ahmed:2023rky,Antusch:2023zjk,Maji:2024cwv,Antusch:2024nqg,Chitose:2024pmz}. These could evade the constraints if they decay early enough or could explain the signal if they decay around the corresponding time, leading to a GW spectrum that increases with frequency that could match the shape of the signal. We also note a recent analysis by the LIGO--VIRGO--KAGRA collaborations \cite{LIGOScientific:2025kry}, which obtain bounds from searches at higher frequencies.
 For the network model of Ref.~\cite{Ringeval:2005kr} this analysis excludes $G_{\rm N}\mu\gtrsim 10^{-14}$. In contrast, assuming the model of Ref.~\cite{Blanco-Pillado:2013qja} (which we adopt for our GW signal prediction following \cite{Buchmuller:2021mbb}) the corresponding bound is weakened to $G_{\rm N}\mu\gtrsim 10^{-8}$. In addition, there exist scenarios in which string networks may evade these high-frequency constraints altogether \cite{Datta:2024bqp,Hu:2025sxv}.

In Figure~\ref{fig:GWspectrum} we plot the GW spectrum produced by metastable cosmic string networks with a fixed $\kappa=485$ for the dominant breaking mechanisms in thermal and inflationary production. For thermal production we show the results assuming only finite-temperature breaking and including the (more important but less certain) effect of pre-existing monopoles. For inflationary production, monopoles produced during inflation dominate over late-time breaking. In the two production scenarios, we assume that the reheating temperature $T_{\rm RH}=v$ and $H_I/(2\pi)=v$ respectively; for larger values the GW spectrum cuts off at higher frequencies.  We consider only emission from loops, as described by Eq.~\eqref{eq:fullspectrum}. The sensitivities of current (LIGO, PTA) and future (LISA, ET) GW detectors, together with the SGWB signal observed by PTA are also indicated. 

Meanwhile, Figure~\ref{fig:frequencies} shows the dependence of the low-frequency cutoff $f_{\rm low}$ on $\kappa$, with the predictions considering each separate breaking mechanism shown individually. These results are analogous to Figure~\ref{fig:Tdall}, but translated to $f_{\rm low}$ to connect to observational implications. As expected, pre-existing monopoles lead to the earliest breaking, followed by early-time finite-temperature corrections to the breaking rate, and lastly breaking at late times due to quantum tunnelling. Fitting the PTA signal requires $f_{\rm low} \sim 10^{-8} \, {\rm Hz}$, corresponding to $\kappa \sim \mathcal{O}(10^2) - \mathcal{O}(10^3)$, much larger than the commonly assumed value $\kappa \simeq 64$. For the pre-existing monopoles and inflationary production, we again assume $T_{\rm RH}=v$ and $H_I/(2\pi)=v$ respectively. For the non-uniform tunnelling, we take $x_{\rm s} = 1.73$ and evaluate $F(x_{\rm s})$ using the distribution in Eq. \eqref{eq:tailmu}, as discussed at the end of Section \ref{sec:late}.

For $\kappa \lesssim 100$, the cutoff $f_{\rm low}$ lies well above the LIGO sensitivity band (see Figure~\ref{fig:frequencies}), implying that any signal would appear only at high frequencies. A metastable-string network destroyed by any of the processes we consider may therefore generate a GW signal peaked at high frequencies \textit{without} requiring an unreasonably small scale separation ($\kappa \lesssim 30$), where the thin-string approximation breaks down. High frequency gravitational waves are largely unconstrained; the only substantial bound on the string tension in this band comes from limits on additional relativistic degrees of freedom, i.e. $N_{\rm eff}$, at BBN, which implies $G_{\rm N}\mu \lesssim 10^{-4}$. The potential observational prospects for frequencies above the LIGO band are summarised in the recent review \cite{Aggarwal:2025noe}, while a discussion specifically on high-frequency gravitational waves from cosmic strings can be found in \cite{Servant:2023tua}.

\section{Hidden sector QCD-like theories} \label{sec:qcdlike}

As mentioned in the Introduction, the flux tubes of a confining non-Abelian hidden sector can behave like cosmic strings \cite{Yamada:2022aax,PureYM22} and are, in many cases, metastable. For definiteness, we consider an $\mathrm{SU}(N_{\rm DC})$ gauge theory with fundamental quarks. If the lightest quark has mass $m_{\mathcal Q} \gg \Lambda_{\rm DC}$, where $\Lambda_{\rm DC}$ is the dark-sector confinement scale, then the flux tubes are metastable and have a tension that is parametrically $\mu_{\rm DC} \sim \Lambda_{\rm DC}^{2}$. Since flux tubes can end on quarks, these objects play broadly the same role as monopoles do for local strings. Moreover, if the hidden-sector reheating temperature exceeds $\Lambda_{\rm DC}$, a network of flux tubes is expected to form as the temperature drops below $\Lambda_{\rm DC}$ and the theory confines.

At late times, flux tube breaking proceeds through an instanton process similar to that described in Section~\ref{sec:stringbreaking}, with a zero-temperature, uniform-tension breaking rate of
\beq \label{eq:quarkbreaking}
\Gamma \simeq \frac{\mu_{\rm DC}}{2\pi} e^{-\pi \kappa_{\rm DC}}~, \quad \kappa_{\rm DC} \equiv \frac{m_{\mathcal{Q}}^2}{\mu_{\rm DC}}~,
\eeq
analogous to Eq.~\eqref{eq:4}. In a realistic network, the late-time breaking rate will be modified relative to Eq.~\eqref{eq:quarkbreaking} by inhomogeneities in the tension. Moreover, as in the gauge-string case, the breaking rate is parametrically enhanced at temperatures near the confinement scale. As in the analysis of Section~\ref{sec:reheat}, this early-time breaking is expected to be more important than late-time breaking for the eventual destruction of the network. 
Additionally, the presence of heavy quarks at the time the network forms will prevent the existence of infinite flux tubes. Heavy quarks can be produced through early-universe mechanisms similar to those responsible for monopole production. During inflation, gravitational production occurs at a rate schematically similar to that for monopoles; see Ref.~\cite{Chung:2011ck} for a recent analysis.\footnote{Additionally, efficient production can also occur around the end of inflation, even if $m_{\mathcal Q} \gg H_I$, provided that $m_{\mathcal Q}$ is smaller than the inflaton mass \cite{Ema:2018ucl}.} Assuming a reheating temperature $T_{\rm RH} < m_{\mathcal Q}$, thermal production is Boltzmann suppressed; however, for a network produced by thermal effects, the resulting quark abundance is typically sufficient to play an important role in the network’s eventual destruction.

We note, however, that flux tubes are qualitatively different from Abelian Higgs strings, for example in their classification. In the Abelian Higgs model, strings carry an integer winding number, whereas in $\mathrm{SU}(N_{\rm DC})$ gauge theories flux tubes are characterised by their “$k$-ality,” $k = 0,1,\ldots,N_{\rm DC}-1$, which labels the transformation of the flux under the $\mathbb{Z}_{N_{\rm DC}}$ centre symmetry. Combined with the non-perturbative nature of flux tubes, this leads to a range of interesting possible dynamics. For example, two $k=1$ (fundamental) strings can, when oriented appropriately, overlap and form a $k=2$ string. Further discussion can be found in Ref.~\cite{PureYM22,Nakamura:2025ptd}, and we will carry out a detailed analysis of the structure of flux-tube networks in future work.

\section{Discussion} \label{sec:discuss}

In this paper, we studied the formation, evolution, and (under plausible assumptions) the eventual destruction of a network of metastable cosmic strings. We considered networks produced by both thermal and inflationary fluctuations, and argued that in each case there are physical processes that prevent the formation of truly infinite strings, leaving only finite segments and smaller loops. Using a simple model, we analytically argued that interactions among these finite segments do not lead to infinite strings forming. For large hierarchies between the monopole mass and string tension ($\kappa\gg 1$), the resulting segments are nevertheless extremely super-horizon in length and behave effectively like a network of infinite strings until their endpoints re-enter each other’s horizons and decay begins. This typically occurs much earlier than would be expected by quantum tunnelling breaking alone. The main phenomenological implication is that, for the string formation scenarios we have considered, much larger values of $\kappa$ are required for the network to decay at observationally relevant times. This can be viewed positively for model building, since it relaxes the degree of tuning required in the reheating temperature (or inflationary Hubble scale) to achieve a string-forming transition without a monopole-forming transition as well.  Theories with relatively small $\kappa$, for which the network decays very early, could generate a GW spectrum peaked at high frequencies; see, for example, \cite{Aggarwal:2025noe} for a recent review of observational efforts in this range.

A key assumption, and the main uncertainty, in our analysis of a network's destruction time is that  string loops do not re-percolate to form effectively infinite strings. This seems plausible because, as noted earlier, most of the string length is initially in finite segments  (though a significant fraction may later be transferred into loops as the network approaches scaling). The impact of loops is an interesting open question that we leave for future study; the results presented here are sufficient to establish early-time breaking as potentially crucial, and to show that the evolution of a metastable network is more complex than previously assumed.  A natural way to investigate the loop dynamics would be to perform Nambu-Goto simulations with a realistic initial distribution of string lengths.  Field-theoretic simulations, analogous to those used in Section~\ref{sec:late} but including also the field giving rise to monopoles, might also be useful. Interesting related studies have been carried out in the context of semilocal strings \cite{Achucarro:2005vpt,Nunes:2011sf,Achucarro:2013mga}, where a network with long strings does reform. However, in that case the monopoles are global and experience a distance-independent force, which likely leads to dynamics that are very different from the local monopoles we consider. A useful review is provided in Ref.~\cite{Kibble:2015twa}, while other related work appears in Refs.~\cite{Vilenkin:1982hm,COPELAND1988445,Martins:2009hj}. Friction on the strings \cite{Martins:1995tg}, or monopoles \cite{Bais:1981zm}, due to the thermal bath might also play an important role in the dynamics.

In the case of thermal production of strings, we also made the crucial assumption 
that reheating and thermalisation are instantaneous. As discussed in Section~\ref{sec:reheat}, this is unlikely to hold in realistic cosmological histories. It would therefore be interesting to study the evolution of metastable string networks under more realistic assumptions, also including, e.g., the effects of spatial inhomogeneities in the inflaton's decay products. An important question is whether the symmetry associated with the string-forming transition can be restored without producing so many monopoles at earlier epochs (e.g., prior to full thermalisation) that the resulting string segments are short compared to the horizon.

We have focused on the simplest realisations of metastable strings and assumed the absence of small parameters that could separate the monopole mass from the symmetry breaking scale. In more complex theories, e.g. \cite{Ng:2008mp}, a wider range of dynamics might be possible, and our conclusions about which breaking mechanisms dominate could change (another interesting scenario is if a monopole population is partly diluted by inflation \cite{Lazarides:2022jgr}).
It would likewise be interesting to consider global strings, for which most of the energy is delocalised from the core and the breaking dynamics may differ significantly. 
Another realisation of metastable strings, arising from a single-stage symmetry breaking, was recently discussed in \cite{Ingoldby:2025wcl}. Although such models do not feature a pre-existing monopole population, we expect that our arguments regarding finite-temperature corrections to the breaking rate and late-time breaking should still apply.  
Cosmic fundamental string theory strings may also be metastable, although this depends sensitively on the details of the underlying string theory and the compactification. Finally, as briefly discussed in Section~\ref{sec:qcdlike}, hidden-sector Yang-Mills flux tubes provide a particularly natural realisation of metastable strings.
We leave the task of exploring the dynamics across the full landscape of theories giving rise to metastable strings as another direction for future work.

We also reanalysed string breaking at late times due to quantum tunnelling. Our results from simulations are sufficient to establish that breaking this way primarily occurs at rare high-tension locations; however, there are many directions in which the analysis could be improved. For example, a theoretical understanding of the distribution of string tensions, possibly even explaining the functional form of the fit in Eq.~\eqref{eq:tailmu} would be very useful and would make our extrapolation from simulations to the physical regime more robust. This might also allow the impact of backreaction from gravitational wave emission to be estimated. Nambu-Goto simulations might also be able to provide insight. Relatedly, we expect that there might be some cutoff in the tension distribution. If this could be determined reliably, our conservative lower bound on the breaking rate could be converted into a prediction.

To calculate the GW spectrum produced as the network is destroyed, we have followed the approach and approximations of  Ref.~\cite{Buchmuller:2021mbb}. In this, the emission from the motion of the monopole endpoints is negligible (see also \cite{Martin:1996ea,Leblond:2009fq}). Nevertheless, further detailed analysis would be valuable. 
For example, Ref.~\cite{Chitose:2025qyt} recently suggested that collisions between monopoles and thermal fluctuations on the strings can prevent the monopole-antimonopole pair from oscillating, thereby suppressing the signal. Observational signatures associated with the monopole dynamics may also be possible, especially if the monopoles carry Standard Model charges, which would also be interesting to explore in future work. Additionally, the fact that monopoles are present throughout the evolution of the network (albeit at an average density of less than one per Hubble patch until the strings are destroyed), could plausibly affect the resulting signals.

\section*{Acknowledgements}
We thank John March-Russell for very useful discussions and collaboration on related work and Valerie Domcke for insightful comments on a draft. EH and LT thank Marco Gorghetto, Anton Sokolov and Giovanni Villadoro for collaboration on related work, and EH thanks Ed Copeland for a useful discussion. LT is supported by the Science and Technology Facilities Council Doctoral Training Partnership under the grant ST/Y509474/1. EC is supported by the Clarendon Fund Scholarship in partnership with the Oxford-Berman Graduate Scholarship. EH acknowledges the UK Research and Innovation Future Leader Fellowship MR/V024566/1. NKvIJ expresses gratitude for the financial support from Merton College, Oxford, through the Buckee Scholarship, and from the Clarendon Fund through the Clarendon Fund Scholarship. For the purpose of open access, the author has applied a CC BY public copyright license to any Author Accepted Manuscript (AAM) version arising from this submission.

\appendix

\section{Toy model of string evolution} \label{app:toymodel}
 
We consider a simple model for the interaction of long (super-horizon) but finite string segments. This neglects Hubble expansion and the destruction of string length that is required to maintain scaling (e.g. by the production of sub-horizon string loops). However, it is sufficient to argue that interactions between long string segments do not lead to infinite strings forming.

Consider a set of $N\gg 1$ positive real numbers $\{l_i\,|\,i=1\ldots N \}$ with mean $\mu$. Two distinct indices $i$ and $j$ are chosen independently with probability proportional to the corresponding $l$ (i.e. $l_i$ and $l_j$ respectively). We then replace these values by
\beq \label{eq:updaterule}
\begin{aligned}
l_i' &=\epsilon_1 l_i + \epsilon_2 l_j \\
l_j' &=(1-\epsilon_1) l_i + (1-\epsilon_2) l_j~.
\end{aligned}
\eeq
where $\epsilon_{1,2}$ are random variables drawn from the uniform distribution on $[0,1]$. 

The $l_i$ represent the lengths of long strings in a system, and the update rule corresponds to the process of two long strings intersecting and recombining at random points along their lengths.
The probability that $l_i$ is selected for update is proportional to its value to reflect the fact that longer strings will interact more frequently; in a physical setting roughly one interaction per Hubble length per Hubble time is expected. 
This model can be viewed as a mean-field analogue of a kinetic process: pairs of string segments “collide” and redistribute their lengths stochastically.

\begin{remark}
Under repeated application of the update rule, the unique stationary distribution of string lengths is
\beq
f(l)= \frac{1}{\mu} e^{-l/\mu}~.
\eeq
\end{remark}

\begin{proof}
We denote expectation values by $E[\ldots]$, with $E_{X,Y}[\ldots]$ indicating that the expectation value is over the random variables $X$ and $Y$ etc.

We call the probability distribution function (pdf) for the string length $f(l)$, which has mean
\beq
\mu=\int_0^\infty l f(l) \,dl~.
\eeq
The Laplace transform of $f$ is
\beq
\Phi(s) \equiv E_L[e^{-s L}]=\int_0^\infty e^{-sl}f(l)\,dl~.
\eeq
Given the rule above, the pdf for a string of length $l$ to be selected for update is
\beq
g(l)= \frac{l}{\mu} f(l)~.
\eeq
It is useful to define the function 
\beq
G(s)=\int_0^\infty e^{-sl}g(l)\, dl~, 
\eeq
which satisfies
\beq \label{eq:Gident}
G(s)=\frac{1}{\mu}\int_0^\infty l e^{-sl}f(l)\,dl= -\frac{1}{\mu}\Phi'(s) = \frac{-1}{\mu} \partial_s \int_0^\infty e^{-sl}f(l) \,dl~.
\eeq
On a stationary distribution $G(s)$ satisfies
\beq \label{eq:update}
G(s) = E_{\epsilon_1,L_1,\epsilon_2,L_2}[e^{-s( \epsilon_1 L_1+\epsilon_2 L_2)}] =\left( E_{\epsilon_1,L}[e^{-s \epsilon_1 L}] \right)^2~,
\eeq
where the second equality is valid in the large $N$ limit. Eq.~\eqref{eq:update} expresses the condition that the length distribution is invariant under the update: the left-hand side is the (Laplace transform) of the distribution of string length selected to be updated, and the right-hand side is the (Laplace transform) of the distribution of new strings after the update.

The right-hand side of Eq.~\eqref{eq:update} can be written as
\beq
\begin{aligned}
\Phi(s) &= \left( \int_0^1d\epsilon_1 \int_0^\infty dl\, \frac{l}{\mu} f(l)e^{-s \epsilon_1 l} \right)^2\\
&= \left( \int_0^1 d\epsilon_1 \, G(s\epsilon_1) \right)^2~.
\end{aligned}
\eeq
Using Eq.~\eqref{eq:Gident}, we have
\beq
\int_0^1d\epsilon_1 \, G(s\epsilon_1) = \frac{1}{s}\int_0^s du\, G(u) = \frac{1}{s}\left( \frac{1-\Phi(s)}{\mu} \right)~,
\eeq
and hence on the stationary distribution 
\beq
-\frac{1}{\mu}\Phi'(s) = \left(\frac{1-\Phi(s)}{\mu s} \right)^2~.
\eeq
Solving this differential equation, we find
\beq
\Phi(s)= \frac{1+ (C-1)s \mu}{1+Cs \mu} ~,
\eeq
where $C$ is a constant of integration. Hence, taking the inverse Laplace transform,
\beq
f(l)=\frac{1-C}{C} \delta(l) + \frac{1}{\mu C^2}e^{-l/(C \mu)}~.
\eeq
We identify $1-1/C$ as the fraction of string length in zero-length strings in the initial condition, which are never selected for update. Taking the physical case of $C=1$ leads to the  claimed result.

\end{proof}

This shows that the frequency of string segments of length exceeding the mean is exponentially suppressed. Numerical experiments suggest that this stationary distribution is an attractor of the dynamics, which is physically reasonable since strings longer than the mean tend to shorten upon interaction.

We note that, in addition to not capturing the loss of string length by loop formation, this model cannot account for loops being absorbed onto long strings, which would increase the mean string length. Additionally, it cannot be used as a model of interactions among loops, which can merge or split since they carry no topological charge.

A slightly more realistic model for the evolution of string segments can be obtained as follows. We begin with some probability distribution function of (finite) string lengths, normalised to the initial Hubble length. During each successive Hubble time, we assume a number of interactions between long strings such that (on average) each Hubble length of string interacts once. The update rule in each case is the same as Eq.~\eqref{eq:updaterule}. Additionally, we discard any strings that are smaller than the Hubble distance (i.e. sub-horizon strings that are expected to disappear). 

Although too complex to be analytically tractable, it is straightforward to implement numerically. We have done so starting from an exponential distribution, which is a reasonable initial condition given Sections~\ref{sec:reheat} and \ref{sec:inflation}. We find that, as expected, the strings are all destroyed once the initial mean length is similar to the Hubble length, consistent with what is expected from the simpler model in which small strings are not discarded. In particular, strings that are much larger than the mean initial length are exponentially rare.

\section{Additional results from simulations}\label{app:furthersim}

\begin{figure}[t]
    \centering
    \begin{subfigure}[t]{0.49\textwidth}
         \centering
         \includegraphics[width=\textwidth]{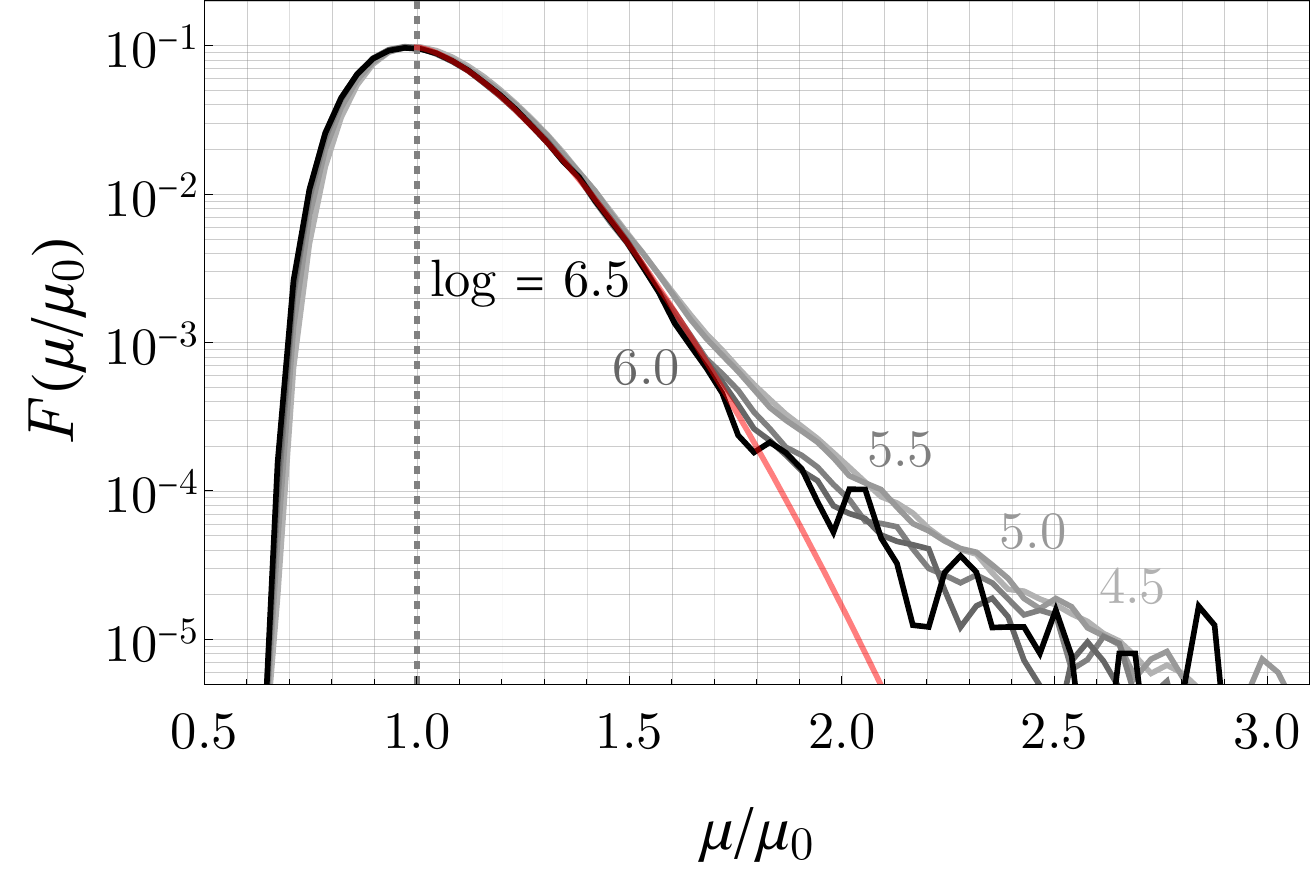}
    \end{subfigure}
    \hfill
    \begin{subfigure}[t]{0.49\textwidth}
          \centering
         \includegraphics[width=\textwidth]{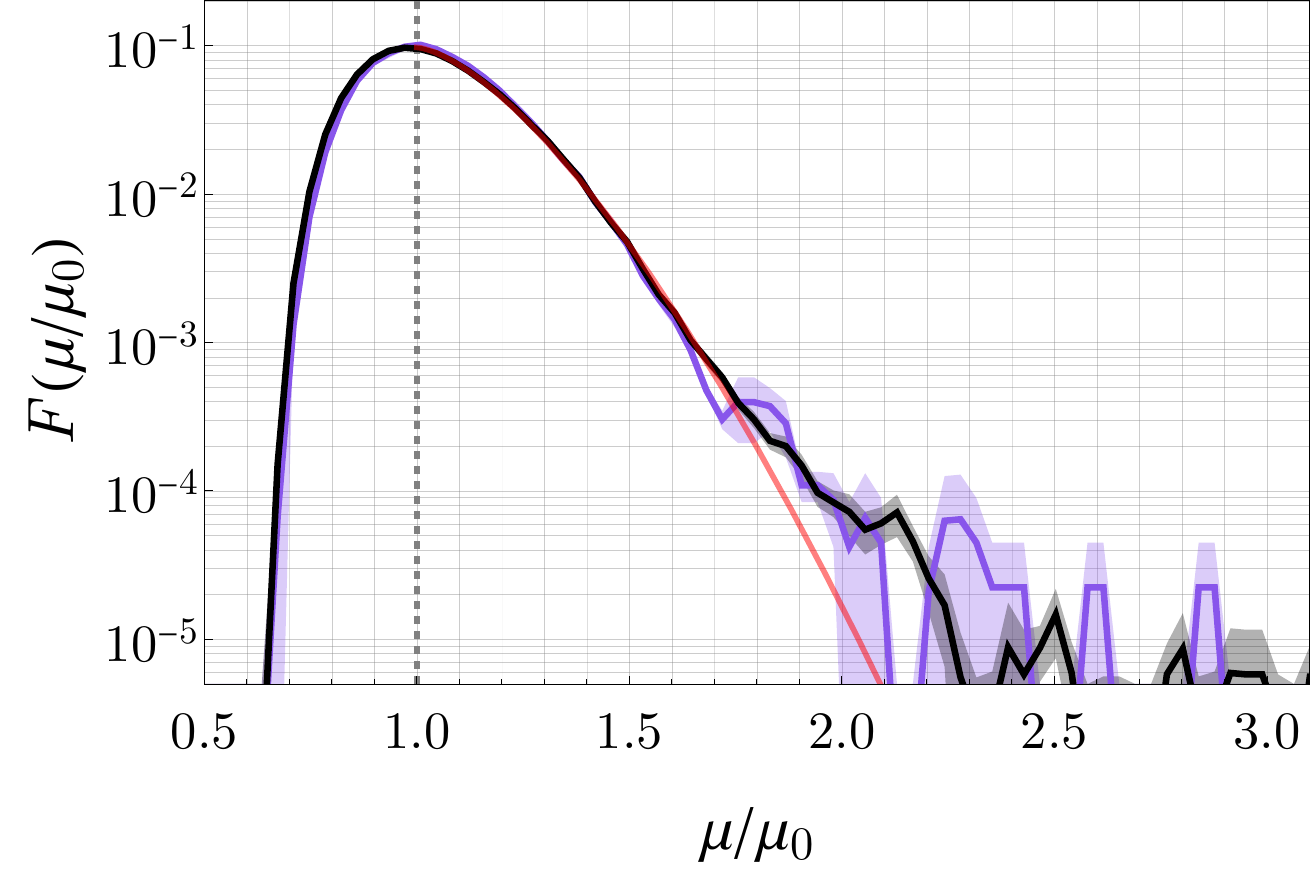}
    \end{subfigure}
    \hfill
    \caption{\textbf{Left:} Distribution of tensions for an Abelian-Higgs critical string network (with straight string tension $\mu_0 = 2\pi v^2$) extracted from numerical simulations at different simulation times, expressed in terms of $\log = \log(m_r/H)$. The red line corresponds to the late-time extrapolation, given in Eq.~\eqref{eq:tailmu}. \textbf{Right:} Distribution of tensions at $\log(m_r/H) = 6.3$, for simulations with 2 core points (black line) and 3 core points (purple line). The two distributions are almost indistinguishable for $x < 1.7$, and diverge only in the (initial-conditions dependent) high-tension tail.}
    \label{fig:mudistappendix}
\end{figure}

To study the dynamics of a string network and obtain the local tension distribution, we use numerical simulations of critically coupled Abelian-Higgs strings. In particular, we numerically integrate the classical field theory defined by the Abelian-Higgs model at critical coupling on an expanding FLRW background in radiation domination. Given suitable initial conditions, strings will naturally form within the simulation volume at the start of the simulation. Here we discuss additional details of the simulation setup used and present some results concerning the tension distribution. We refer to \cite{usfuture} for further details and a detailed study of systematic uncertainties. 

The simulation volume is taken to be a cube with comoving side length $\tilde{L}$, discretised on a lattice of size $N^3$ with lattice spacing $\Delta = \tilde{L}/N$ and periodic boundary conditions. For the main runs used to extract the tension data presented in the main text, we set $N = 3072$. At each conformal time $t$, the discretised complex scalar field $\phi_{\bm x}$ is defined as being equal to its continuum counterpart at each lattice point $\bm{x}$, while the gauge connections $\theta_{\bm x}^i = e A_i({\bm x}) \Delta$ are defined along the links in the positive $i$th direction from the point $\bm x$. The scalar and vector fields are evolved in equal conformal timesteps $\Delta_t$, where good convergence requires $\Delta_t < \Delta /3$. For the evolution, we employ a fourth order leapfrog algorithm based on \cite{Hindmarsh:2014rka, Klaer:2017qhr}, which has discretisation errors vanishing as $\mathcal{O}(\Delta^4)$. 

On this grid, we set up our initial conditions using a pre-evolution method. The scalar field is initialized with a random value at each lattice point and smoothed using a low-pass filter. The network thus obtained has very high density, defined as length of string per Hubble patch, and a large amount of energy in scalar and gauge radiation. The fields then undergo an unphysical, overdamped evolution on an inflationary background which  dilutes the string network and the radiation, similarly to the method described in \cite{Klaer:2017qhr}. When the density of the network reaches some target value, the pre-evolution ends and the field configuration is handed over to the main evolution algorithm. 

String cores are identified by measuring the gauge-invariant winding through each lattice plaquette, defined in \cite{Kajantie:1998bg}. The number of lattice points inside the string core $n_r = ( \Delta m_r)^{-1}$ determines the resolution of our simulations; for local strings, taking $n_r \gtrsim 2$ is sufficient to approximate the continuum limit. 
The local tension $\mu$ is computed using the method described in Section~\ref{sec:late} of the main text, see in particular Eq.~\eqref{eq:tensionmethod}. The distribution $F(x,t)$ is obtained by binning  $\mu$ with bin size $\delta \mu/\mu_0 = 0.2$. 

The distribution in Figure~\ref{fig:mudist} is obtained by averaging over 5 simulations with identical initial conditions. The distribution is peaked around $x = 1$, with a tail at high tensions which is the main focus of our discussion in Section~\ref{sec:late}. 

In the left panel of Figure~\ref{fig:mudistappendix}, we show the distribution $F(x,t)$ at different simulation times. We observe a more prominent high-tension tail at early times, which we interpret as a leftover transient from the initial conditions or due to frequent string intersections at early times, gradually evolving towards the behaviour in Eq.~\eqref{eq:tailmu}.

The right panel of Figure~\ref{fig:mudistappendix} compares the tension distribution at late simulation times for batches of simulations with $n_r = 2, 3$, i.e. lower and higher resolutions. We see that the shape of the distribution is consistent with Eq.~\eqref{eq:tailmu} regardless of resolution.  Therefore, we conclude that the results presented in Section~\ref{sec:late} are free from systematic uncertainties from finite lattice spacing, and in particular the existence of a high tension tail in $F(x,t)$ is unlikely to be a simulation artifact.

\section{Temperature dependence of the tunnelling action}\label{app:tunnelling}
\begin{figure}
    \centering
    \includegraphics[width=0.7\linewidth]{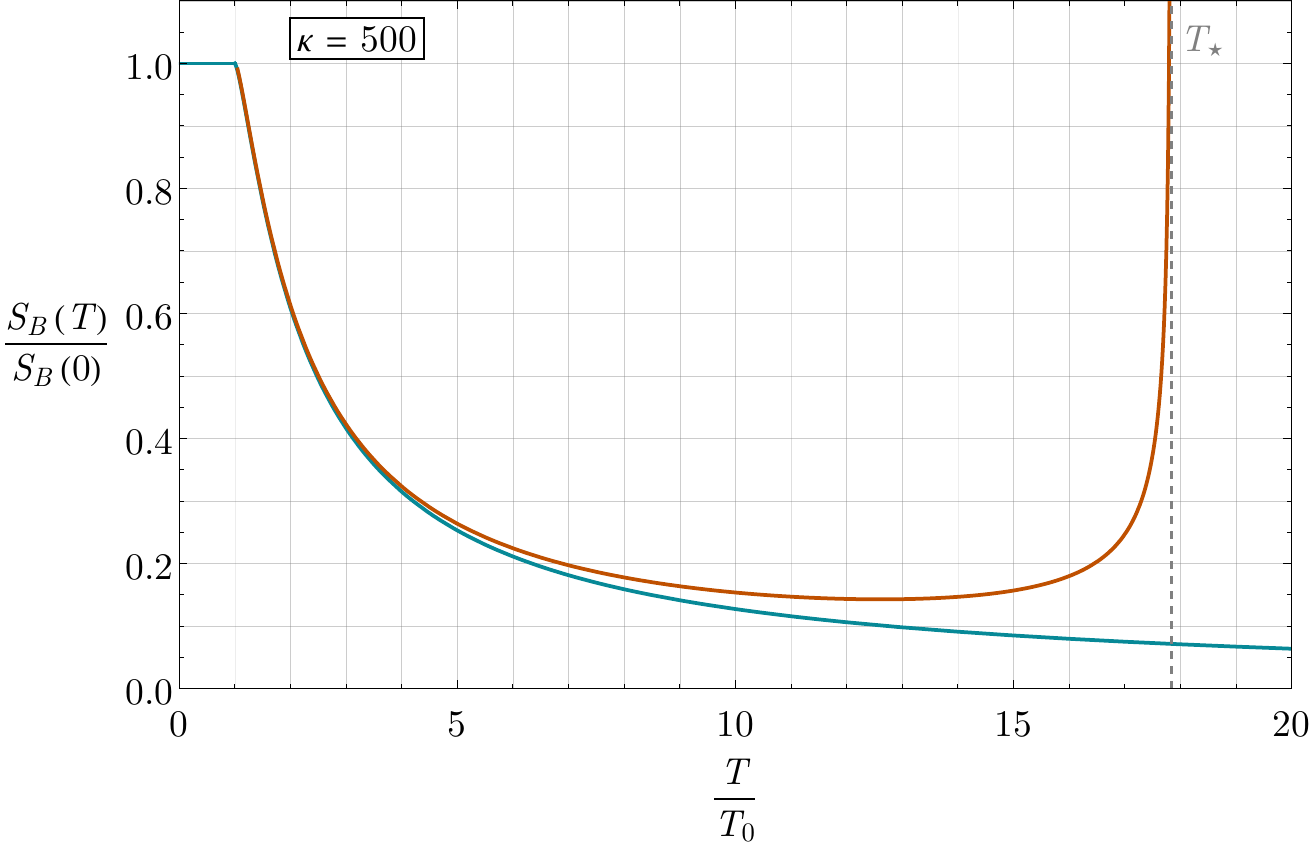}
    \caption{Ratio of the temperature dependent bounce action $S_B (T)$ over its zero-temperature value, for $\kappa = 500$. The dashed vertical line corresponds to the critical temperature $T_\star$ at which the string network forms. The blue line corresponds to the form of the action used in the main text, with $v = {\rm const.}$; the red line to a temperature dependent $v$ with the form of Eq.~\eqref{eq:vT}.}
    \label{fig:actionapp}
\end{figure}
In Section~\ref{sec:reheat} of the main text, we found the temperature dependence of the tunnelling action $S_B(T)$, which controls the nucleation rate of monopole-antimonopole pairs along the string. The resulting expression, given in Eq.~\eqref{eq:SBT}, neglects the possible temperature dependence of the vacuum expectation value $v$, and hence of the string tension $\mu \sim v^2$. Indeed, for a typical second-order phase transition we expect that 
\begin{equation}\label{eq:vT}
    v(T) =  v\sqrt{1-(T/T_\star)^2} \, ,
\end{equation}
where $T_\star$ is the critical temperature. Including this dependence alters the form of the tunnelling action close to $T_\star$, such that it becomes impossible to break the strings while they form. However, when $\kappa \gg 1$ the value of $S_B$ quickly decreases to match Eq.~\eqref{eq:SBT} for temperatures just below $T_\star$, such that the breaking rate is still exponentially faster compared to Eq.~\eqref{eq:4}. This is shown in Figure~\ref{fig:actionapp}. We see that the smallest value of $S_B$ is no longer at $T = T_\star$, but rather at some lower temperature $T = T_{\rm min} \sim T_\star/\mathcal{O}(1)$. The analysis of Section~\ref{subsec:finiteT} would then be unaffected, except for the replacement $T_\star \to T_{\rm min}$ in Eq.~\eqref{eq:elli}. This in turn modifies our estimates for $\kappa$ by some $\mathcal{O}(1)$ number, which approaches $1$ when $\kappa$ is large. 

\bibliographystyle{JHEP}
\bibliography{biblio}

\end{document}